\input harvmac

\input amssym
\let\includefigures=\iftrue

\input rotate
\input epsf

\def\hat{\widehat}

\def\CZ{{\cal Z}}

\noblackbox

\lref\DavisQE{
  J.~L.~Davis, F.~Larsen and N.~Seiberg,
  ``Heterotic strings in two dimensions and new stringy phase transitions,''
  arXiv:hep-th/0505081.
}

\lref\Davisup{
  J.~L.~Davis, to appear.
}

\lref\KlebanovQA{
   I.~R.~Klebanov,
   ``String theory in two-dimensions,''
   arXiv:hep-th/9108019.
}

\lref\GinspargIS{
   P.~H.~Ginsparg and G.~W.~Moore,
   ``Lectures on 2-D gravity and 2-D string theory,''
   arXiv:hep-th/9304011.
}

\lref\McGuiganQP{
  M.~D.~McGuigan, C.~R.~Nappi and S.~A.~Yost,
  ``Charged black holes in two-dimensional string theory,''
  Nucl.\ Phys.\ B {\bf 375}, 421 (1992)
  [arXiv:hep-th/9111038].
}

\lref\StromingerTM{
  A.~Strominger,
  ``A matrix model for AdS(2),''
  JHEP {\bf 0403}, 066 (2004)
  [arXiv:hep-th/0312194].
}

\lref\WittenIM{
  E.~Witten,
  ``Bound states of strings and p-branes,''
  Nucl.\ Phys.\ B {\bf 460}, 335 (1996)
  [arXiv:hep-th/9510135].
}

\lref\PolchinskiSM{
  J.~Polchinski and A.~Strominger,
  ``New Vacua for Type II String Theory,''
  Phys.\ Lett.\ B {\bf 388}, 736 (1996)
  [arXiv:hep-th/9510227].
}

\lref\OSV{
  H.~Ooguri, A.~Strominger and C.~Vafa,
  ``Black hole attractors and the topological string,''
  Phys.\ Rev.\ D {\bf 70}, 106007 (2004)
  [arXiv:hep-th/0405146].
}

\lref\HorowitzJC{
  G.~T.~Horowitz and J.~Polchinski,
  ``Self gravitating fundamental strings,''
  Phys.\ Rev.\ D {\bf 57}, 2557 (1998)
  [arXiv:hep-th/9707170].
}

\lref\AtickSI{
  J.~J.~Atick and E.~Witten,
  ``The Hagedorn Transition And The Number Of Degrees Of Freedom Of String
  Theory,''
  Nucl.\ Phys.\ B {\bf 310}, 291 (1988).
}

\lref\HananyIE{
  A.~Hanany and E.~Witten,
  ``Type IIB superstrings, BPS monopoles, and three-dimensional gauge
  dynamics,''
  Nucl.\ Phys.\ B {\bf 492}, 152 (1997)
  [arXiv:hep-th/9611230].
}

\lref\KapustinHI{
  A.~Kapustin,
  ``Noncritical superstrings in a Ramond-Ramond background,''
  JHEP {\bf 0406}, 024 (2004)
  [arXiv:hep-th/0308119].
}

\lref\DeWolfeQF{
  O.~DeWolfe, R.~Roiban, M.~Spradlin, A.~Volovich and J.~Walcher,
  ``On the S-matrix of type 0 string theory,''
  JHEP {\bf 0311}, 012 (2003)
  [arXiv:hep-th/0309148].
}

\lref\BerkovitsTG{
  N.~Berkovits, S.~Gukov and B.~C.~Vallilo,
  ``Superstrings in 2D backgrounds with R-R flux and new extremal black
  Nucl.\ Phys.\ B {\bf 614}, 195 (2001)
  [arXiv:hep-th/0107140].
}

\lref\KlebanovWG{
  I.~R.~Klebanov, J.~Maldacena and N.~Seiberg,
  ``Unitary and complex matrix models as 1-d type 0 strings,''
  Commun.\ Math.\ Phys.\  {\bf 252}, 275 (2004)
  [arXiv:hep-th/0309168].
}

\lref\GrossZZ{
  D.~J.~Gross and J.~Walcher,
  ``Non-perturbative RR potentials in the c(hat) = 1 matrix model,''
  JHEP {\bf 0406}, 043 (2004)
  [arXiv:hep-th/0312021].
}

\lref\SeibergNM{
  N.~Seiberg and D.~Shih,
  ``Branes, rings and matrix models in minimal (super)string theory,''
  JHEP {\bf 0402}, 021 (2004)
  [arXiv:hep-th/0312170].
}

\lref\StromingerTM{
  A.~Strominger,
  ``A matrix model for AdS(2),''
  JHEP {\bf 0403}, 066 (2004)
  [arXiv:hep-th/0312194].
}

\lref\GaiottoGD{
  D.~Gaiotto,
  ``Long strings condensation and FZZT branes,''
  arXiv:hep-th/0503215.
}

 \lref\FriessTQ{
  J.~J.~Friess and H.~Verlinde,
  ``Hawking effect in 2-D string theory,''
  arXiv:hep-th/0411100.
}

\lref\DanielssonYI{
  U.~H.~Danielsson,
  ``A matrix model black hole: Act II,''
  JHEP {\bf 0402}, 067 (2004)
  [arXiv:hep-th/0312203].
}

\lref\GukovYP{
  S.~Gukov, T.~Takayanagi and N.~Toumbas,
  ``Flux backgrounds in 2D string theory,''
  JHEP {\bf 0403}, 017 (2004)
  [arXiv:hep-th/0312208].
}

\lref\DavisXB{
  J.~L.~Davis, L.~A.~Pando Zayas and D.~Vaman,
  ``On black hole thermodynamics of 2-D type 0A,''
  JHEP {\bf 0403}, 007 (2004)
  [arXiv:hep-th/0402152].
}

\lref\DanielssonXF{
  U.~H.~Danielsson, J.~P.~Gregory, M.~E.~Olsson, P.~Rajan and M.~Vonk,
  ``Type 0A 2D black hole thermodynamics and the deformed matrix model,''
  JHEP {\bf 0404}, 065 (2004)
  [arXiv:hep-th/0402192].
}

\lref\TakayanagiJZ{
  T.~Takayanagi,
  ``Notes on D-branes in 2D type 0 string theory,''
  JHEP {\bf 0405}, 063 (2004)
  [arXiv:hep-th/0402196].
}

\lref\VerlindeGT{
  H.~Verlinde,
  ``Superstrings on AdS(2) and superconformal matrix quantum mechanics,''
  arXiv:hep-th/0403024.
}

\lref\ParkYC{
  J.~Park and T.~Suyama,
  ``Type 0A matrix model of black hole, integrability and holography,''
  Phys.\ Rev.\ D {\bf 71}, 086002 (2005)
  [arXiv:hep-th/0411006].
}

\lref\SeibergEI{
  N.~Seiberg and D.~Shih,
  ``Flux vacua and branes of the minimal superstring,''
  JHEP {\bf 0501}, 055 (2005)
  [arXiv:hep-th/0412315].
}

\lref\SeibergBX{
  N.~Seiberg,
  ``Observations on the moduli space of two dimensional string theory,''
  JHEP {\bf 0503}, 010 (2005)
  [arXiv:hep-th/0502156].
}

\lref\nonsinglets{
  J.~Maldacena,
   ``Long strings in two dimensional string theory and non-singlets in the
   matrix model,''
  arXiv:hep-th/0503112.
}
\lref\DouglasUP{
  M.~R.~Douglas, I.~R.~Klebanov, D.~Kutasov, J.~Maldacena, E.~Martinec and N.~Seiberg,
  ``A new hat for the c = 1 matrix model,''
  arXiv:hep-th/0307195.
}

\lref\KlebanovWG{
  I.~R.~Klebanov, J.~Maldacena and N.~Seiberg,
  ``Unitary and complex matrix models as 1-d type 0 strings,''
  Commun.\ Math.\ Phys.\  {\bf 252}, 275 (2004)
  [arXiv:hep-th/0309168].
}

\lref\wittenbaryon{
  E.~Witten,
  ``Baryons and branes in anti de Sitter space,''
  JHEP {\bf 9807}, 006 (1998)
  [arXiv:hep-th/9805112].
}

\lref\MorrisBW{ T.~R.~Morris, ``2-D Quantum Gravity, Multicritical
Matter And Complex Matrices,'' FERMILAB-PUB-90-136-T
}

\lref\LapanQZ{
  J.~M.~Lapan and W.~Li,
  ``Falling D0-branes in 2D superstring theory,''
  arXiv:hep-th/0501054.
}

\lref\DavisXI{
  J.~L.~Davis and R.~McNees,
  `Boundary counterterms and the thermodynamics of 2-D black holes,''
  arXiv:hep-th/0411121.
}

%

\lref\MartinecQT{
  E.~Martinec and K.~Okuyama,
  ``Scattered results in 2D string theory,''
  JHEP {\bf 0410}, 065 (2004)
  [arXiv:hep-th/0407136].
}

 \lref\HoQP{
  P.~M.~Ho,
  ``Isometry of AdS(2) and the c = 1 matrix model,''
  JHEP {\bf 0405}, 008 (2004)
  [arXiv:hep-th/0401167].
}

\lref\GrossHE{ D.~J.~Gross and E.~Witten, ``Possible Third Order
Phase Transition In The Large N Lattice Gauge Theory,'' Phys.\
Rev.\ D {\bf 21}, 446 (1980).
}

\lref\DiFrancescoRU{ P.~Di Francesco, ``Rectangular Matrix Models
and Combinatorics of Colored Graphs,'' Nucl.\ Phys.\ B {\bf 648},
461 (2003) [arXiv:cond-mat/0208037].
}

\lref\MartinecHT{ E.~J.~Martinec, G.~W.~Moore and N.~Seiberg,
``Boundary operators in 2-D gravity,'' Phys.\ Lett.\ B {\bf 263},
190 (1991).
}

\lref\minahan{ J.~A.~Minahan, ``Matrix models with boundary terms
and the generalized Painleve II equation,'' Phys.\ Lett.\ B {\bf
268}, 29 (1991);
J.~A.~Minahan, ``Schwinger-Dyson equations for unitary matrix
models with boundaries,'' Phys.\ Lett.\ B {\bf 265}, 382 (1991).
}

\lref\PeriwalGF{ V.~Periwal and D.~Shevitz, ``Unitary Matrix
Models As Exactly Solvable String Theories,'' Phys.\ Rev.\ Lett.\
{\bf 64}, 1326 (1990);
V.~Periwal and D.~Shevitz, ``Exactly Solvable Unitary Matrix
Models: Multicritical Potentials And Correlations,'' Nucl.\ Phys.\
B {\bf 344}, 731 (1990).
}

\lref\PeetWN{
  A.~W.~Peet and J.~Polchinski,
  ``UV/IR relations in AdS dynamics,''
  Phys.\ Rev.\ D {\bf 59}, 065011 (1999)
  [arXiv:hep-th/9809022].
}

\lref\AharonyTI{
  O.~Aharony, S.~S.~Gubser, J.~M.~Maldacena, H.~Ooguri and Y.~Oz,
  ``Large N field theories, string theory and gravity,''
  Phys.\ Rept.\  {\bf 323}, 183 (2000)
  [arXiv:hep-th/9905111].
}

\lref\BerkoozCQ{
  M.~Berkooz, M.~Rozali and N.~Seiberg,
  ``Matrix description of M theory on T**4 and T**5,''
  Phys.\ Lett.\ B {\bf 408}, 105 (1997)
  [arXiv:hep-th/9704089].
}

\lref\AharonyXN{
  O.~Aharony, A.~Giveon and D.~Kutasov,
  ``LSZ in LST,''
  Nucl.\ Phys.\ B {\bf 691}, 3 (2004)
  [arXiv:hep-th/0404016].
}

\lref\AharonyKS{
  O.~Aharony,
  ``A brief review of 'little string theories',''
  Class.\ Quant.\ Grav.\  {\bf 17}, 929 (2000)
  [arXiv:hep-th/9911147].
}

\lref\SeibergZK{
  N.~Seiberg,
  ``New theories in six dimensions and matrix description of M-theory on  T**5
  and T**5/Z(2),''
  Phys.\ Lett.\ B {\bf 408}, 98 (1997)
  [arXiv:hep-th/9705221].
}

\lref\AharonyUB{
  O.~Aharony, M.~Berkooz, D.~Kutasov and N.~Seiberg,
  ``Linear dilatons, NS5-branes and holography,''
  JHEP {\bf 9810}, 004 (1998)
  [arXiv:hep-th/9808149].
}

\lref\DouglasUP{ M.~R.~Douglas, I.~R.~Klebanov, D.~Kutasov,
J.~Maldacena, E.~Martinec and N.~Seiberg, ``A new hat for the c =
1 matrix model,'' arXiv:hep-th/0307195.
}
\lref\AlexandrovFH{
  S.~Y.~Alexandrov, V.~A.~Kazakov and I.~K.~Kostov,
  ``Time-dependent backgrounds of 2D string theory,''
  Nucl.\ Phys.\ B {\bf 640}, 119 (2002)
  [arXiv:hep-th/0205079].
}

\lref\KostovTK{
  I.~K.~Kostov,
  ``Integrable flows in c = 1 string theory,''
  J.\ Phys.\ A {\bf 36}, 3153 (2003)
  [Annales Henri Poincare {\bf 4}, S825 (2003)]
  [arXiv:hep-th/0208034].
}

\lref\AlexandrovPZ{
  S.~Y.~Alexandrov and V.~A.~Kazakov,
  ``Thermodynamics of 2D string theory,''
  JHEP {\bf 0301}, 078 (2003)
  [arXiv:hep-th/0210251].
}

\lref\AlexandrovQK{
  S.~Y.~Alexandrov, V.~A.~Kazakov and I.~K.~Kostov,
  ``2D string theory as normal matrix model,''
  Nucl.\ Phys.\ B {\bf 667}, 90 (2003)
  [arXiv:hep-th/0302106].
}

\lref\YinIV{
  X.~Yin,
  ``Matrix models, integrable structures, and T-duality of type 0
string
  theory,''
  Nucl.\ Phys.\ B {\bf 714}, 137 (2005)
  [arXiv:hep-th/0312236].
}

\lref\SethiES{
  S.~Sethi, C.~Vafa and E.~Witten,
  ``Constraints on low-dimensional string compactifications,''
  Nucl.\ Phys.\ B {\bf 480}, 213 (1996)
  [arXiv:hep-th/9606122].
}

\lref\AlexandrovKS{
  S.~Alexandrov,
  ``D-branes and complex curves in c=1 string theory,''
  JHEP {\bf 0405}, 025 (2004)
  [arXiv:hep-th/0403116].
}

\lref\AlexandrovCG{
  S.~Y.~Alexandrov and I.~K.~Kostov,
  ``Time-dependent backgrounds of 2D string theory:
Non-perturbative effects,''
  JHEP {\bf 0502}, 023 (2005)
  [arXiv:hep-th/0412223].
}

\lref\TeschnerRD{
  J.~Teschner,
  ``On Tachyon condensation and open-closed duality in the c = 1
string
  theory,''
  arXiv:hep-th/0504043.
}

\lref\CrnkovicMS{ C.~Crnkovic, M.~R.~Douglas and G.~W.~Moore,
``Physical Solutions For Unitary Matrix Models,'' Nucl.\ Phys.\ B
{\bf 360}, 507 (1991).
}

\lref\DiFrancescoXZ{ P.~Di Francesco, H.~Saleur and J.~B.~Zuber,
``Generalized Coulomb Gas Formalism For Two-Dimensional Critical
Models Based On SU(2) Coset Construction,'' Nucl.\ Phys.\ B {\bf
300}, 393 (1988).
}

\lref\DijkgraafPP{ R.~Dijkgraaf, S.~Gukov, V.~A.~Kazakov and
C.~Vafa, ``Perturbative analysis of gauged matrix models,''
arXiv:hep-th/0210238.
}

\lref\AganagicQJ{
  M.~Aganagic, R.~Dijkgraaf, A.~Klemm, M.~Marino and C.~Vafa,
  ``Topological strings and integrable hierarchies,''
  arXiv:hep-th/0312085.
}

\lref\Fukuda{ T.~Fukuda and K.~Hosomichi, ``Super Liouville theory
with boundary,'' Nucl.\ Phys.\ B {\bf 635}, 215 (2002)
[arXiv:hep-th/0202032].
}

\lref\PolchinskiZF{
  J.~Polchinski,
  ``Evaluation Of The One Loop String Path Integral,''
  Commun.\ Math.\ Phys.\  {\bf 104}, 37 (1986).
}

\lref\KostovXW{ I.~K.~Kostov, ``Solvable statistical models on a
random lattice,'' Nucl.\ Phys.\ Proc.\ Suppl.\  {\bf 45A}, 13
(1996) [arXiv:hep-th/9509124].
}

\lref\Hollowood{ T.~J.~Hollowood, L.~Miramontes, A.~Pasquinucci
and C.~Nappi, ``Hermitian versus anti-Hermitian one matrix models
and their hierarchies,'' Nucl.\ Phys.\ B {\bf 373}, 247 (1992)
[arXiv:hep-th/9109046].
}

\lref\NappiBI{ C.~R.~Nappi, ``Painleve-II And Odd Polynomials,''
Mod.\ Phys.\ Lett.\ A {\bf 5}, 2773 (1990).
}

\lref\MartinecKA{ E.~J.~Martinec, ``The annular report on
non-critical string theory,'' arXiv:hep-th/0305148.
}

\lref\AlexandrovNN{ S.~Y.~Alexandrov, V.~A.~Kazakov and
D.~Kutasov, ``Non-Perturbative Effects in Matrix Models and
D-branes,'' arXiv:hep-th/0306177.
}

\lref\DalleyBR{ S.~Dalley, C.~V.~Johnson, T.~R.~Morris and
A.~Watterstam, ``Unitary matrix models and 2-D quantum gravity,''
Mod.\ Phys.\ Lett.\ A {\bf 7}, 2753 (1992) [arXiv:hep-th/9206060].
}

\lref\johnsonflows{ C.~V.~Johnson, T.~R.~Morris and A.~Watterstam,
``Global KdV flows and stable 2-D quantum gravity,'' Phys.\ Lett.\
B {\bf 291}, 11 (1992) [arXiv:hep-th/9205056].
}

\lref\DalleyQG{ S.~Dalley, C.~V.~Johnson and T.~Morris,
``Multicritical complex matrix models and nonperturbative 2-D
quantum gravity,'' Nucl.\ Phys.\ B {\bf 368}, 625 (1992).
}

\lref\LafranceWY{ R.~Lafrance and R.~C.~Myers, ``Flows For
Rectangular Matrix Models,'' Mod.\ Phys.\ Lett.\ A {\bf 9}, 101
(1994) [arXiv:hep-th/9308113].
}

\lref\SeibergEB{ N.~Seiberg, ``Notes On Quantum Liouville Theory
And Quantum Gravity,'' Prog.\ Theor.\ Phys.\ Suppl.\  {\bf 102},
319 (1990).
}

\lref\gelfand{ I.~M.~Gelfand and L.~A.~Dikii, ``Asymptotic
Behavior Of The Resolvent Of Sturm-Liouville Equations And The
Algebra Of The Korteweg-De Vries Equations,'' Russ.\ Math.\
Surveys {\bf 30}, 77 (1975) [Usp.\ Mat.\ Nauk {\bf 30}, 67
(1975)].
}

\lref\BrowerMN{ R.~C.~Brower, N.~Deo, S.~Jain and C.~I.~Tan,
``Symmetry breaking in the double well Hermitian matrix models,''
Nucl.\ Phys.\ B {\bf 405}, 166 (1993) [arXiv:hep-th/9212127].
}

\lref\CrnkovicWD{ C.~Crnkovic, M.~R.~Douglas and G.~W.~Moore,
``Loop equations and the topological phase of multi-cut matrix
models,'' Int.\ J.\ Mod.\ Phys.\ A {\bf 7}, 7693 (1992)
[arXiv:hep-th/9108014].
}

\lref\BerKleb{ M.~Bershadsky and I.~R.~Klebanov, ``Genus One Path
Integral In Two-Dimensional Quantum Gravity,'' Phys.\ Rev.\ Lett.\
{\bf 65}, 3088 (1990).
} \lref\BerKlebnew{ M.~Bershadsky and I.~R.~Klebanov, ``Partition
functions and physical states in two-dimensional quantum gravity
and supergravity,'' Nucl.\ Phys.\ B {\bf 360}, 559 (1991).
} \lref\igor{ I.~R.~Klebanov, ``String theory in two-dimensions,''
arXiv:hep-th/9108019.
}

\lref\ginsparg{ P.~Ginsparg and G.~W.~Moore, ``Lectures On 2-D
Gravity And 2-D String Theory,'' arXiv:hep-th/9304011.
}

\lref\DineXK{
  M.~Dine, N.~Seiberg and E.~Witten,
  ``Fayet-Iliopoulos Terms In String Theory,''
  Nucl.\ Phys.\ B {\bf 289}, 589 (1987).
}

\lref\WittenDG{
  E.~Witten,
  ``Some Properties Of O(32) Superstrings,''
  Phys.\ Lett.\ B {\bf 149}, 351 (1984).
}

\lref\DiFrancescoGinsparg{P.~Di Francesco, P.~Ginsparg and
J.~Zinn-Justin,``2-D Gravity and random matrices,'' Phys.\ Rept.\
{\bf 254}, 1 (1995) [arXiv:hep-th/9306153].
}

\lref\joe{J.~Polchinski, ``What is String Theory?''
arXiv:hep-th/9411028
}

\lref\mgv{ J.~McGreevy and H.~Verlinde, ``Strings from tachyons:
The $c = 1$ matrix reloaded,'' arXiv:hep-th/0304224.
}

\lref\SchomerusVV{ V.~Schomerus, ``Rolling tachyons from Liouville
theory,'' arXiv:hep-th/0306026.
}

\lref\GaiottoYF{ D.~Gaiotto, N.~Itzhaki and L.~Rastelli, ``On the
BCFT description of holes in the c = 1 matrix model,''
arXiv:hep-th/0307221.
}

\lref\GutperleIJ{ M.~Gutperle and P.~Kraus, ``D-brane dynamics in
the c = 1 matrix model,'' arXiv:hep-th/0308047.
}

\lref\KapustinHI{ A.~Kapustin, ``Noncritical superstrings in a
Ramond-Ramond background,'' arXiv:hep-th/0308119.
}

\lref\GiveonWN{ A.~Giveon, A.~Konechny, A.~Pakman and A.~Sever,
``Type 0 strings in a 2-d black hole,'' arXiv:hep-th/0309056.
}

\lref\DixonIZ{
   L.~J.~Dixon and J.~A.~Harvey,
   ``String Theories In Ten-Dimensions Without Space-Time
Supersymmetry,''
   Nucl.\ Phys.\ B {\bf 274}, 93 (1986).
}

\lref\GiveonZZ{
 A.~Giveon and A.~Sever,
 ``Strings in a 2-d extremal black hole,''
 JHEP {\bf 0502}, 065 (2005)
 [arXiv:hep-th/0412294].
}

\lref\KutasovZM{
 D.~Kutasov, E.~J.~Martinec and M.~O'Loughlin,
 ``Vacua of M-theory and N=2 strings,''
 Nucl.\ Phys.\ B {\bf 477}, 675 (1996)
 [arXiv:hep-th/9603116].
}

\lref\KutasovVH{
 D.~Kutasov and E.~J.~Martinec,
 ``M-branes and N = 2 strings,''
 Class.\ Quant.\ Grav.\  {\bf 14}, 2483 (1997)
 [arXiv:hep-th/9612102].
}

\lref\OsorioHI{
  M.~A.~R.~Osorio and M.~A.~Vazquez-Mozo,
  Phys.\ Lett.\ B {\bf 280}, 21 (1992)
  [arXiv:hep-th/9201044].
}

\lref\OsorioVG{
  M.~A.~R.~Osorio and M.~A.~Vazquez-Mozo,
  Phys.\ Rev.\ D {\bf 47}, 3411 (1993)
  [arXiv:hep-th/9207002].
}

\lref\SeibergBY{
   N.~Seiberg and E.~Witten,
   ``Spin Structures In String Theory,''
   Nucl.\ Phys.\ B {\bf 276}, 272 (1986).
}

\lref\KutasovUA{
   D.~Kutasov and N.~Seiberg,
   ``Noncritical Superstrings,''
   Phys.\ Lett.\ B {\bf 251}, 67 (1990).
}

\lref\BouwknegtVA{ P.~Bouwknegt, J.~G.~McCarthy and K.~Pilch,
`BRST analysis of physical states for 2-D (super)gravity coupled
to (super)conformal matter,'' arXiv:hep-th/9110031.
}

\lref\PandaGE{ S.~Panda and S.~Roy, ``BRST cohomology ring in
$c(M) < 1$ NSR string theory,'' Phys.\ Lett.\ B {\bf 358}, 229
(1995) [arXiv:hep-th/9507054].
}

\lref\ItohIX{ K.~Itoh and N.~Ohta, ``Spectrum of two-dimensional
(super)gravity,'' Prog.\ Theor.\ Phys.\ Suppl.\  {\bf 110}, 97
(1992) [arXiv:hep-th/9201034].
}

\lref\BouwknegtAM{ P.~Bouwknegt, J.~G.~McCarthy and K.~Pilch,
``Ground ring for the 2-D NSR string,'' Nucl.\ Phys.\ B {\bf 377},
541 (1992) [arXiv:hep-th/9112036].
}

\lref\ItohIY{ K.~Itoh and N.~Ohta, ``BRST cohomology and physical
states in 2-D supergravity coupled to $c \le 1$ matter,'' Nucl.\
Phys.\ B {\bf 377}, 113 (1992) [arXiv:hep-th/9110013].
}

\lref\ImbimboIA{
  C.~Imbimbo, S.~Mahapatra and S.~Mukhi,
  ``Construction of physical states of nontrivial ghost number in c < 1
string
  theory,''
  Nucl.\ Phys.\ B {\bf 375}, 399 (1992).
}

\lref\LianGK{ B.~H.~Lian and G.~J.~Zuckerman, ``New Selection
Rules And Physical States In 2-D Gravity: Conformal Gauge,''
Phys.\ Lett.\ B {\bf 254}, 417 (1991).
}

\lref\WittenZD{
   E.~Witten,
   ``Ground ring of two-dimensional string theory,''
   Nucl.\ Phys.\ B {\bf 373}, 187 (1992)
   [arXiv:hep-th/9108004].
}

\lref\AdamsRB{
   A.~Adams, X.~Liu, J.~McGreevy, A.~Saltman and E.~Silverstein,
   ``Things fall apart: Topology change from winding tachyons'',
   arXiv:hep-th/0502021.
}

\lref\AtickSI{
   J.~J.~Atick and E.~Witten,
   ``The Hagedorn Transition And The Number Of Degrees Of Freedom Of
String
   Theory,''
   Nucl.\ Phys.\ B {\bf 310}, 291 (1988).
}

\lref\AlvarezUK{
   E.~Alvarez and M.~A.~R.~Osorio,
   ``Cosmological Constant Versus Free Energy For Heterotic Strings,''
   Nucl.\ Phys.\ B {\bf 304}, 327 (1988)
   [Erratum-ibid.\ B {\bf 309}, 220 (1988)].
}

\lref\DineVU{
  M.~Dine, P.~Y.~Huet and N.~Seiberg,
  ``Large And Small Radius In String Theory,''
  Nucl.\ Phys.\ B {\bf 322}, 301 (1989).
}

\lref\MaloneyRR{
   A.~Maloney, E.~Silverstein and A.~Strominger,
   ``De Sitter space in noncritical string theory,''
   [arXiv:hep-th/0205316].
}

\lref\McGreevyKB{
   J.~McGreevy and H.~Verlinde,
   ``Strings from tachyons: The c = 1 matrix reloaded,''
   JHEP {\bf 0312}, 054 (2003)
   [arXiv:hep-th/0304224].
}

\lref\KlebanovKM{
   I.~R.~Klebanov, J.~Maldacena and N.~Seiberg,
   ``D-brane decay in two-dimensional string theory,''
   JHEP {\bf 0307}, 045 (2003)
   [arXiv:hep-th/0305159].
}

\lref\TakayanagiSM{
   T.~Takayanagi and N.~Toumbas,
``A matrix model dual of type 0B string theory in two
dimensions,''
   JHEP {\bf 0307}, 064 (2003)
   [arXiv:hep-th/0307083].
}

\lref\GukovYP{
   S.~Gukov, T.~Takayanagi and N.~Toumbas,
``Flux backgrounds in 2D string theory,''
   JHEP {\bf 0403}, 017 (2004)
   [arXiv:hep-th/0312208].
}

\lref\McGreevyDN{ J.~McGreevy, S.~Murthy and H.~Verlinde,
``Two-dimensional superstrings and the supersymmetric matrix
model,''
   JHEP {\bf 0404}, 015 (2004)
   [arXiv:hep-th/0308105].
}

\lref\TakayanagiGE{ T.~Takayanagi, ``Comments on 2D type IIA
string and matrix model,'' JHEP {\bf 0411}, 030 (2004)
[arXiv:hep-th/0408086].
}

\lref\GinspargBX{
   P.~H.~Ginsparg,
   ``Comment On Toroidal Compactification Of Heterotic Superstrings,''
   Phys.\ Rev.\ D {\bf 35}, 648 (1987).
}

\lref\AlvarezGaumeJB{
   L.~Alvarez-Gaume, P.~H.~Ginsparg, G.~W.~Moore and C.~Vafa,
   ``An O(16) X O(16) Heterotic String,''
   Phys.\ Lett.\ B {\bf 171}, 155 (1986).
}

\lref\McGuiganQP{
   M.~D.~McGuigan, C.~R.~Nappi and S.~A.~Yost,
   ``Charged black holes in two-dimensional string theory,''
   Nucl.\ Phys.\ B {\bf 375}, 421 (1992)
   [arXiv:hep-th/9111038].
}

\lref\HarveyNA{
   J.~A.~Harvey, D.~Kutasov and E.~J.~Martinec,
  ``On the relevance of tachyons,''
   [arXiv:hep-th/0003101].
}

\lref\LustTJ{
   D.~Lust and S.~Theisen,
   ``Lectures On String Theory,''
   Lect.\ Notes Phys.\  {\bf 346}, 1 (1989).
}

\lref\PolchinskiRR{
   J.~Polchinski,
   ``String theory. Vol. 2: Superstring theory and beyond,''
   Cambridge University Press (1998).
}

\lref\PolchinskiZF{
   J.~Polchinski,
   ``Evaluation Of The One Loop String Path Integral,''
   Commun.\ Math.\ Phys.\  {\bf 104}, 37 (1986).
}

\lref\GrossUB{
   D.~J.~Gross and I.~R.~Klebanov,
   ``One-Dimensional String Theory On A Circle,''
   Nucl.\ Phys.\ B {\bf 344}, 475 (1990).
}
\lref\DouglasUP{
   M.~R.~Douglas, I.~R.~Klebanov, D.~Kutasov, J.~Maldacena, E.~Martinec
and N.~Seiberg,
   ``A new hat for the c = 1 matrix model,''
   [arXiv:hep-th/0307195].
}

\lref\SeibergBX{
   N.~Seiberg,
   ``Observations on the moduli space of two dimensional string theory,''
   [arXiv:hep-th/0502156].
}

\lref\DixonAC{
   L.~J.~Dixon, P.~H.~Ginsparg and J.~A.~Harvey,
   ``(Central Charge C) = 1 Superconformal Field Theory,''
   Nucl.\ Phys.\ B {\bf 306}, 470 (1988).
}

\lref\AlvarezSJ{
   E.~Alvarez and M.~A.~R.~Osorio,
   ``Superstrings At Finite Temperature,''
   Phys.\ Rev.\ D {\bf 36}, 1175 (1987).
}

\lref\BrienPN{
   K.~H.~O'Brien and C.~I.~Tan,
   ``Modular Invariance Of Thermopartition Function And Global Phase
Structure
   Of Heterotic String,''
   Phys.\ Rev.\ D {\bf 36}, 1184 (1987).
}

\lref\SeibergEB{
  N.~Seiberg,
  ``Notes On Quantum Liouville Theory And Quantum Gravity,''
  Prog.\ Theor.\ Phys.\ Suppl.\  {\bf 102}, 319 (1990).
}

\lref\KlebanovQA{
   I.~R.~Klebanov,
   ``String theory in two-dimensions,''
   arXiv:hep-th/9108019.
}

\lref\GinspargIS{
   P.~H.~Ginsparg and G.~W.~Moore,
   ``Lectures on 2-D gravity and 2-D string theory,''
   arXiv:hep-th/9304011.
}

\lref\BergmanYP{
   O.~Bergman and S.~Hirano,
   ``The cap in the hat: Unoriented 2D strings and matrix(-vector)
models,''
   JHEP {\bf 0401}, 043 (2004)
   [arXiv:hep-th/0311068].
}

\lref\GomisVI{
   J.~Gomis and A.~Kapustin,
   ``Two-dimensional unoriented strings and matrix models,''
   JHEP {\bf 0406}, 002 (2004)
   [arXiv:hep-th/0310195].
}

\lref\Harv{ J.L.~Karczmarek and A.~Strominger, ``Matrix
cosmology,'' arXiv:hep-th/0309138.
}

\lref\Kitp{O.~DeWolfe, R.~Roiban, M.~Spradlin, A.~Volovich and
J.~Walcher, ``On the S-matrix of type 0 string theory,''
arXiv:hep-th/0309148.
}

\lref\KlebanovKM{ I.~R.~Klebanov, J.~Maldacena and N.~Seiberg,
``D-brane decay in two-dimensional string theory,''
arXiv:hep-th/0305159.
}

\lref\KutasovUA{
  D.~Kutasov and N.~Seiberg,
  ``Noncritical Superstrings,''
  Phys.\ Lett.\ B {\bf 251}, 67 (1990).
}

\lref\McGreevyEP{ J.~McGreevy, J.~Teschner and H.~Verlinde,
``Classical and quantum D-branes in 2D string theory,''
arXiv:hep-th/0305194.
}

\lref\TeschnerQK{ J.~Teschner, ``On boundary perturbations in
Liouville theory and brane dynamics in noncritical string
theories,'' arXiv:hep-th/0308140.
}

\lref\zz{ A.~B.~Zamolodchikov and A.~B.~Zamolodchikov, ``Liouville
field theory on a pseudosphere,'' arXiv:hep-th/0101152.
}

\lref\KazakovCH{ V.~A.~Kazakov and A.~A.~Migdal, ``Recent Progress
In The Theory Of Noncritical Strings,'' Nucl.\ Phys.\ B {\bf 311},
171 (1988).
}

\lref\TakayanagiSM{ T.~Takayanagi and N.~Toumbas, ``A Matrix Model
Dual of Type 0B String Theory in Two Dimensions,''
arXiv:hep-th/0307083.
}

\lref\SW{N. Seiberg and E. Witten, unpublished.}

\lref\GrossAY{ D.~J.~Gross and N.~Miljkovic, ``A Nonperturbative
Solution of $D = 1$ String Theory,'' Phys.\ Lett.\ B {\bf 238},
217 (1990);
%
E.~Brezin, V.~A.~Kazakov and A.~B.~Zamolodchikov, ``Scaling
Violation in a Field Theory of Closed Strings in One Physical
Dimension,'' Nucl.\ Phys.\ B {\bf 338}, 673 (1990);
%
P.~Ginsparg and J.~Zinn-Justin, ``2-D Gravity + 1-D Matter,''
Phys.\ Lett.\ B {\bf 240}, 333 (1990).
} \lref\JevickiQN{ A.~Jevicki, ``Developments in 2-d string
theory,'' arXiv:hep-th/9309115.
}

\lref\DHokerZY{ E.~D'Hoker, ``Classical And Quantal Supersymmetric
Liouville Theory,'' Phys.\ Rev.\ D {\bf 28}, 1346 (1983).
}

\lref\RashkovJX{ R.~C.~Rashkov and M.~Stanishkov, ``Three-point
correlation functions in $N=1$ Super Liouville Theory,'' Phys.\
Lett.\ B {\bf 380}, 49 (1996) [arXiv:hep-th/9602148].
}

\lref\PoghosianDW{ R.~H.~Poghosian, ``Structure constants in the
$N=1$ super-Liouville field theory,'' Nucl.\ Phys.\ B {\bf 496},
451 (1997) [arXiv:hep-th/9607120].
}

\lref\Sennew{ A.~Sen, ``Open-Closed Duality: Lessons from the
Matrix Model,'' arXiv:hep-th/0308068.
}

\lref\DalleyVR{ S.~Dalley, C.~V.~Johnson and T.~Morris,
``Nonperturbative two-dimensional quantum gravity,'' Nucl.\ Phys.\
B {\bf 368}, 655 (1992).
}

\lref\KutasovSV{ D.~Kutasov and N.~Seiberg, ``Number Of Degrees Of
Freedom, Density Of States And Tachyons In String Theory And
Cft,'' Nucl.\ Phys.\ B {\bf 358}, 600 (1991).
}

\lref\GopakumarKI{ R.~Gopakumar and C.~Vafa, ``On the gauge
theory/geometry correspondence,'' Adv.\ Theor.\ Math.\ Phys.\ {\bf
3}, 1415 (1999) [arXiv:hep-th/9811131].
}

\lref\KlebanovHB{ I.~R.~Klebanov and M.~J.~Strassler,
``Supergravity and a confining gauge theory: Duality cascades and
chiSB-resolution of naked singularities,'' JHEP {\bf 0008}, 052
(2000) [arXiv:hep-th/0007191].
}

\lref\MaldacenaYY{ J.~M.~Maldacena and C.~Nunez, ``Towards the
large N limit of pure N = 1 super Yang Mills,'' Phys.\ Rev.\
Lett.\  {\bf 86}, 588 (2001) [arXiv:hep-th/0008001].
}

\lref\VafaWI{ C.~Vafa, ``Superstrings and topological strings at
large N,'' J.\ Math.\ Phys.\  {\bf 42}, 2798 (2001)
[arXiv:hep-th/0008142].
}

\lref\WittenIG{ E.~Witten, ``On The Structure Of The Topological
Phase Of Two-Dimensional Gravity,'' Nucl.\ Phys.\ B {\bf 340}, 281
(1990).
}

\lref\MartinecHT{ E.~J.~Martinec, G.~W.~Moore and N.~Seiberg,
``Boundary operators in 2-D gravity,'' Phys.\ Lett.\ B {\bf 263},
190 (1991).
}

\lref\CachazoJY{ F.~Cachazo, K.~A.~Intriligator and C.~Vafa, ``A
large N duality via a geometric transition,'' Nucl.\ Phys.\ B {\bf
603}, 3 (2001) [arXiv:hep-th/0103067].
}

\lref\BrezinRB{ E.~Brezin and V.~A.~Kazakov, ``Exactly Solvable
Field Theories Of Closed Strings,'' Phys.\ Lett.\ B {\bf 236}, 144
(1990).
}

\lref\DouglasVE{ M.~R.~Douglas and S.~H.~Shenker, ``Strings In
Less Than One-Dimension,'' Nucl.\ Phys.\ B {\bf 335}, 635 (1990).
}

\lref\GrossVS{ D.~J.~Gross and A.~A.~Migdal, ``Nonperturbative
Two-Dimensional Quantum Gravity,'' Phys.\ Rev.\ Lett.\  {\bf 64},
127 (1990).
}

\lref\fzz{V.~Fateev, A.~B.~Zamolodchikov and A.~B.~Zamolodchikov,
``Boundary Liouville field theory. I: Boundary state and boundary
two-point function,'' arXiv:hep-th/0001012.
}

\lref\teschner{ J.~Teschner, ``Remarks on Liouville theory with
boundary,'' arXiv:hep-th/0009138.
}

\lref\AlexandrovFH{
  S.~Y.~Alexandrov, V.~A.~Kazakov and I.~K.~Kostov,
  ``Time-dependent backgrounds of 2D string theory,''
  Nucl.\ Phys.\ B {\bf 640}, 119 (2002)
  [arXiv:hep-th/0205079].
}

\lref\PolchinskiTU{
  J.~Polchinski and Y.~Cai,
  ``Consistency Of Open Superstring Theories,''
  Nucl.\ Phys.\ B {\bf 296}, 91 (1988).
}

\lref\dgjw{  D.~J.~Gross and J.~Walcher,
  ``Non-perturbative RR potentials in the c(hat) = 1 matrix model,''
  JHEP {\bf 0406}, 043 (2004)
  [arXiv:hep-th/0312021].
}

\lref\bruno{
  B.~C.~da Cunha,
  ``Tachyon effective dynamics and de Sitter vacua,''
  Phys.\ Rev.\ D {\bf 70}, 066002 (2004)
  [arXiv:hep-th/0403217].
}

\lref\SeibergBX{
  N.~Seiberg,
  ``Observations on the moduli space of two dimensional string theory,''
  JHEP {\bf 0503}, 010 (2005)
  [arXiv:hep-th/0502156].
}

\lref\DineVU{
  M.~Dine, P.~Y.~Huet and N.~Seiberg,
  ``Large And Small Radius In String Theory,''
  Nucl.\ Phys.\ B {\bf 322}, 301 (1989).
}

\lref\MaldacenaHE{
  J.~Maldacena and N.~Seiberg,
  ``Flux-vacua in two dimensional string theory,''
  JHEP {\bf 0509}, 077 (2005)
  [arXiv:hep-th/0506141].
}

 \lref\MaldacenaHI{
  J.~Maldacena,
  ``Long strings in two dimensional string theory and non-singlets in the
  JHEP {\bf 0509}, 078 (2005)
  [arXiv:hep-th/0503112].
}

\lref\PolchinskiBG{
  J.~Polchinski,
  ``Open heterotic strings,''
  arXiv:hep-th/0510033.
}

\lref\SeibergEB{
  N.~Seiberg,
  ``Notes On Quantum Liouville Theory And Quantum Gravity,''
  Prog.\ Theor.\ Phys.\ Suppl.\  {\bf 102}, 319 (1990).
}

\lref\PolchinskiMH{
  J.~Polchinski,
  ``Remarks On The Liouville Field Theory,''
UTTG-19-90
{\it Presented at Strings '90 Conf., College Station, TX, Mar
12-17, 1990} }

\lref\PolchinskiSM{
  J.~Polchinski and A.~Strominger,
  ``New Vacua for Type II String Theory,''
  Phys.\ Lett.\ B {\bf 388}, 736 (1996)
  [arXiv:hep-th/9510227].
}

\lref\HananyIE{
  A.~Hanany and E.~Witten,
  ``Type IIB superstrings, BPS monopoles, and three-dimensional gauge
  dynamics,''
  Nucl.\ Phys.\ B {\bf 492}, 152 (1997)
  [arXiv:hep-th/9611230].
}

\lref\ItzhakiZR{
  N.~Itzhaki, D.~Kutasov and N.~Seiberg,
  ``Non-supersymmetric deformations of non-critical superstrings,''
  arXiv:hep-th/0510087.
}

\lref\BanksZS{
  T.~Banks, N.~Seiberg and E.~Silverstein,
  ``Zero and one-dimensional probes with N = 8 supersymmetry,''
  Phys.\ Lett.\ B {\bf 401}, 30 (1997)
  [arXiv:hep-th/9703052].
}

\lref\BachasKN{
  C.~P.~Bachas, M.~B.~Green and A.~Schwimmer,
  ``(8,0) quantum mechanics and symmetry enhancement in type I'
  superstrings,''
  JHEP {\bf 9801}, 006 (1998)
  [arXiv:hep-th/9712086].
}

\lref\GomisVI{
  J.~Gomis and A.~Kapustin,
  ``Two-dimensional unoriented strings and matrix models,''
  JHEP {\bf 0406}, 002 (2004)
  [arXiv:hep-th/0310195].
}

\lref\GomisCE{
  J.~Gomis,
  ``Anomaly cancellation in noncritical string theory,''
  arXiv:hep-th/0508132.
}

\def\K3{{\bf K3}}
\def\journal#1&#2(#3){\unskip, \sl #1\ \bf #2 \rm(19#3) }
\def\andjournal#1&#2(#3){\sl #1~\bf #2 \rm (19#3) }

\def\bar{\overline}
\def\hat{\widehat}

\def\tilde{\widetilde}

\def\frac#1#2{{#1\over#2}}

\def\half{\frac12}

\def\inbar{\,\vrule height1.5ex width.4pt depth0pt}
\def\IC{\relax\hbox{$\inbar\kern-.3em{\rm C}$}}
\def\IR{\relax{\rm I\kern-.18em R}}
\def\IP{\relax{\rm I\kern-.18em P}}

%
%

%
\catcode`\@=11
\def\slash#1{\mathord{\mathpalette\c@ncel{#1}}}
\overfullrule=0pt

\def\underrel#1\over#2{\mathrel{\mathop{\kern\z@#1}\limits_{#2}}}

\catcode`\@=12


%

\def \sinh{{\rm sinh}}
\def \cosh{{\rm cosh}}




\rightline{} \Title{ \rightline{hep-th/0511220} }
{\vbox{\centerline{Long Strings, Anomaly Cancellation,}
\centerline{ Phase Transitions, T-duality and Locality}
\centerline{ in the 2d Heterotic String}}}
\medskip

\centerline{\it Nathan Seiberg}
\bigskip
\centerline{School of Natural Sciences}
\centerline{Institute for Advanced Study}
\centerline{Einstein Drive, Princeton, NJ 08540}

\smallskip

\vglue .3cm

\bigskip
\noindent
 We study the noncritical two-dimensional heterotic string.  Long
fundamental strings play a crucial role in the dynamics.  They
cancel anomalies and lead to phase transitions when the system is
compactified on a Euclidean circle. A careful analysis of the
gauge symmetries of the system uncovers new subtleties leading to
modifications of the worldsheet results. The compactification on a
Euclidean thermal circle is particularly interesting.  It leads us
to an incompatibility between T-duality (and its corresponding
gauge symmetry) and locality.

\Date{11/05}

\newsec{Introduction}

Low dimensional theories are useful laboratories for studying
simple phenomena which might have analogs in more physical
dimensions.  In string theory we expect the two-dimensional string
to teach us about more generic phenomena in higher dimensions. The
most studied two-dimensional string theories are the bosonic
theory (for earlier reviews  see, e.g.\
\refs{\KlebanovQA,\GinspargIS}) and its type 0 relatives
\refs{\TakayanagiSM,\DouglasUP}.  These theories have known matrix
model duals, which allow detailed nonperturbative studies. Other
two-dimensional theories include the type II theories
\refs{\SeibergBX,\GukovYP}, the heterotic theories
\refs{\McGuiganQP\GiveonZZ\DavisQE-\Davisup} and various
nonoriented and open string theories
\refs{\GomisVI\BergmanYP-\GomisCE}.

In this note we will continue the investigation of the
two-dimensional heterotic theory\foot{Other theories with somewhat
similar features were studied in
\refs{\OsorioHI\OsorioVG\KutasovZM-\KutasovVH}.}.  We will see
that it shares some of the features of its bosonic, type 0 and
type II counterparts, but it also exhibits many new phenomena.
These include spacetime chirality and non-standard mechanisms for
anomaly cancellation, new subtleties in applying worldsheet
methods and new phase transitions.

As all two-dimensional string theories, our target space has a
linear dilaton along the space direction $\phi$ with the weak
coupling end at $\phi \to -\infty$.  In addition, there is a time
direction $x$ which can have either Lorentzian or Euclidean
signature. We will often regularize the $\phi$ direction in the
weak coupling region and denote its volume by $V$; i.e.\
 \eqn\Vdef{-V< \phi}

The Euclidean time coordinate $x$ can be compactified on a circle
of radius $R$ with various twists.  Then, our target space looks
like an infinite cylinder (see figure 1). It has a noncompact
coordinate $\phi$ and a compact coordinate $x$.

\medskip \ifig\figI{Our target space is an infinite cylinder.
There is a linear dilaton direction $\phi$ and a compact direction
$x$.  We will study it in a Hamiltonian picture in two channels.
The $x$-channel picture describes evolution along $x$ with an
infinite space direction $\phi$.  Here we evaluate a trace.  The
$\phi$-channel picture describes evolution along $\phi$ from the
weak coupling end $\phi \to -\infty$ to the strong coupling end
$\phi \to +\infty$.  Here the space is the circle parameterized by
$x$.}
    {\epsfxsize=0.6\hsize\epsfbox{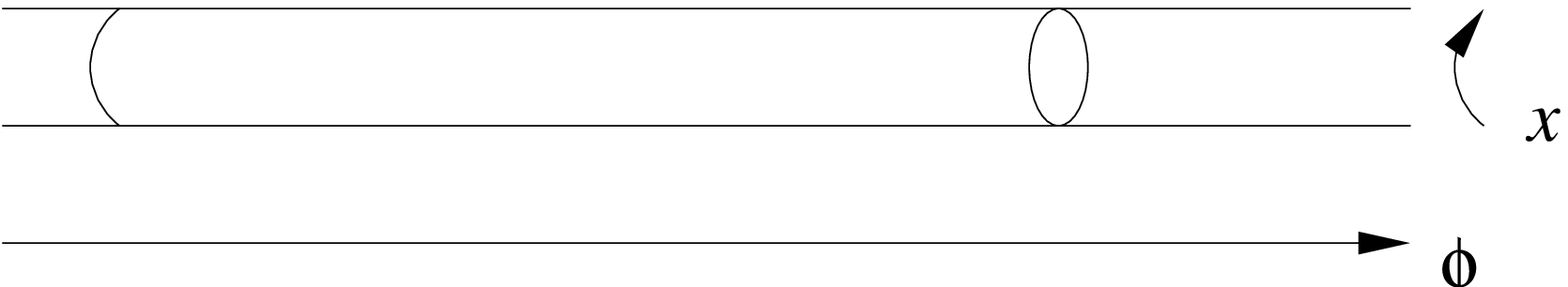}}

In the standard interpretation of this target space $\phi$ is a
space coordinate and $x$ is a Euclidean time coordinate. We will
refer to this interpretation as the $x$-channel.  An alternative
interpretation is to view $\phi$ as Euclidean time and $x$ as a
compact space coordinate.  Here we are interested in the Euclidean
time evolution along $\phi$ from the weak coupling end at $\phi
\to -\infty$ to the strong coupling end.  We will refer to this
picture as the $\phi$-channel (see figure 1).

As in all other two-dimensional noncritical strings, the closed
string spectrum of the theories include only a finite number of
massless particles.  The familiar Hagedorn density of states of
higher dimensional string theory is absent here. Therefore, the
target space theory appears to be similar to an ordinary field
theory with a finite number of fields.  However, in addition to
the closed string spectrum the theory has infinite energy states
associated with infinite long strings, which are stretched along
the entire space direction $\phi$. Such strings were studied in
the two-dimensional bosonic and type 0 theories in
\refs{\MaldacenaHI,\MaldacenaHE}. Long heterotic strings were
recently studied in higher dimensions in \PolchinskiBG.  We will
see that the long heterotic strings have nontrivial degrees of
freedom living on them, and therefore they behave differently than
their bosonic and type 0 relatives.

The long strings couple to the two form field $B$.  Therefore, a
single long string leads to a $B$ tadpole and does not satisfy the
equations of motion.  An infinite long string with the opposite
orientation is a long anti-string and it couples to $B$ with the
opposite sign.  Therefore, a pair of a long string and a long
anti-string carries no $B$ charge and can be added to the system.
Such a pair can annihilate in the interior of the space and form a
folded string which ends at a point $\phi_0$ (see figure 2).  Such
configurations were analyzed in detail in the two-dimensional
bosonic and type 0 theories in \MaldacenaHI.

\medskip \ifig\figII{A pair of an infinite long string and an
infinite long anti-string can be thought of as an infinite string
coming in from the weak coupling end $\phi \to -\infty$, turning
around at $\phi_0$ and moving back to $\phi \to -\infty$.}
    {\epsfxsize=0.6\hsize\epsfbox{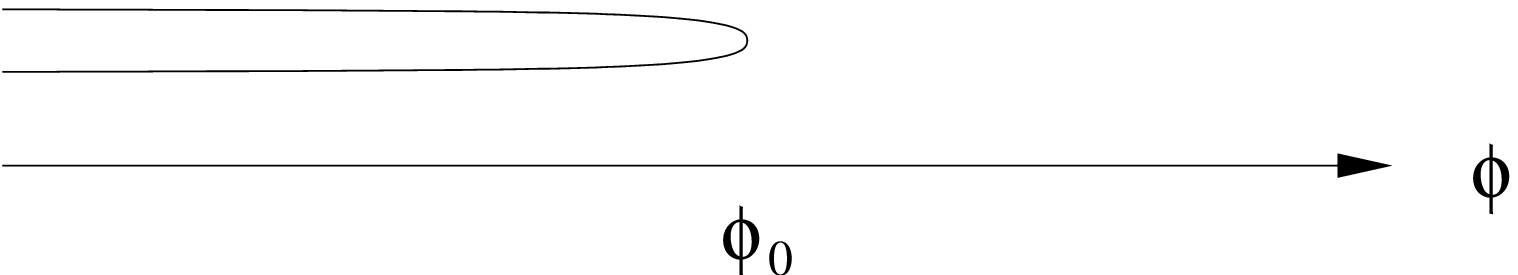}}

We will see that both a single long string and the pairs of
string-anti-string excitations are important to the dynamics of
the heterotic theory.

In section 2 we will review the closed string spectrum of the
three different two-dimensional heterotic string theories in $\Bbb
R^2$: HE, HO and THO.  In section 3 we will quantize the long
string, stressing the differences between its quantization and the
more standard closed string quantization. These long strings will
be used in section 4 to cancel spacetime anomalies.  In section 5
we will start the discussion of compactifications, setting the
notation and making general comments.  Section 6 will introduce
the $\phi$-channel quantization and will explain the subtleties
associated with gauge invariance.  We will see that in some cases
the worldsheet answers need to be modified.

In sections 7 and 8 we will discuss possible phase transitions in
our system.  In section 7 we will consider a transition associated
with a massless complex fermion.  In section 8 we will analyze a
transition associated with a massless complex scalar. Here we will
find a possible discrepancy between the $\phi$-channel and the
$x$-channel pictures.

Sections 9, 10 and 11 will be devoted to specific examples.  In
section 9 we will consider an untwisted circle compactification of
our theories.  In section 10 we will study compactifications
twisted by worldsheet fermion number, and in section 11 we will
discuss thermal compactifications; i.e.\ compactifications twisted
by spacetime fermion number.

We will summarize and will further discuss our results in section
12.

An appendix will examine possible consequences of our results to
the type 0/type II compactifications.

Our general discussion and especially sections 5 - 8, which
precede the concrete examples, might appear somewhat abstract.
Some readers might prefer to read them together with the example
sections 9 - 11.

\newsec{Known heterotic theories}

In addition to the worldsheet fields $x$ and $\phi$, our theory
includes their $(0,1)$ superpartners which are right-moving
fermions $\psi_x$ and $\psi_\phi$, and 24 left-moving fermions
$\lambda^I$. Depending on the sum over spin structures three
theories were identified \refs{\McGuiganQP\DavisQE-\Davisup} (two
of them are analogous to their ten dimensional relatives and hence
the terminology):
 \item{HO} The gauge group is $Spin(24)$ and the
 spectrum includes massless ``tachyons'' in ${\bf 24}$ of
 $Spin(24)$.  Using standard notation their vertex operators are
 \eqn\hov{ T^I(k)=e^{-\varphi}\bar \lambda^Ie^{i kx+
  (1-|k|)\phi} \qquad ;\qquad I=1,...,24}
 \item{HE} The gauge group is $Spin(8) \times E_8$ and the
 spectrum includes massless ``tachyons'' in ${\bf 8_v}$, left-moving
 fermions in ${\bf 8_c}$ and right-moving fermions in ${\bf
 8_s}$ of $Spin(8)$.  Their vertex operators are
   \eqn\hev{\eqalign{
  &T^i(k)=e^{-\varphi}\bar \lambda^i e^{i kx +
  (1-|k|)\phi} \qquad\qquad ;\qquad i=1,...,8 \cr
  &\Psi^\alpha=e^{-\half\varphi+i\half H}\bar S^\alpha
  e^{i kx + (1-|k|)\phi }\qquad ;\qquad
  \alpha =1,...,8 \qquad;\qquad k\geq0 \cr
  &\tilde\Psi^{\dot\alpha}=e^{-\half\varphi-i\half H}\bar
  S^{\dot\alpha} e^{i kx + (1-|k|)\phi}
  \qquad  ;\qquad  \dot\alpha =1,...,8 \qquad;\qquad k\leq0
  }}
 Other versions of this theory which are
 related by $Spin(8)$ triality are physically equivalent.
 \item{THO} This is a twisted (orbifold) version of HO.  The gauge
 group is $Spin(24)$ and the spectrum includes right-moving
 fermions in ${\bf 24}$ of $Spin(24)$
 \eqn\thov{ \tilde \Psi^I(k)=e^{-\half\varphi-i\half
 H} \bar \lambda^Ie^{i kx +
  (1-|k|)\phi} \qquad ;\qquad I=1,...,24 \qquad;\qquad
  k\leq0}
 This theory appears to be anomalous, but we will see below that
 in fact it is consistent. A related
 equivalent theory is obtained from this theory by spacetime parity.

An interesting deformation of the HE and HO theories is a tachyon
background\foot{As is common in the $c=1$ theory, depending on how
we define the theory we might need another factor of $\phi$ in
front of this expression.} $\mu T^1(k=0)=\mu e^{-\varphi}\bar
\lambda^1 e^{\phi}$. This deformation has the effect of breaking
$Spin(8)\to Spin (7)$ and $Spin(24)\to Spin (23)$.  In the zero
ghost number picture the deformation of the (classical) worldsheet
Lagrangian is
 \eqn\mudef{\mu \bar \lambda^1\psi_\phi e^{\phi} + \mu^2 e^{2\phi}}
Here, the second term of order $\mu^2$ arises from imposing
right-moving $(0,1)$ worldsheet supersymmetry\foot{One way to see
that is to consider the two $(0,1)$ superfields $\Phi=\phi+\theta
\psi_\phi$ and $\bar \Lambda^1=\bar \lambda^1 + \theta F^1$, where
$F^1$ is an auxiliary field.  Then the interaction term is $\mu
\int d\theta \bar \Lambda^1 e^\Phi$.  Equation \mudef\ arises
after writing it in components and integrating out the auxiliary
field.}. Interestingly, even though the worldsheet theory has only
$(0,1)$ supersymmetry, the interacting part of this Lagrangian has
$(1,1)$ supersymmetry; it is the well studied $N=1$
super-Liouville theory \refs{\DHokerZY\RashkovJX-\PoghosianDW}.
Borrowing the results of the analysis of this theory, we learn
that it leads to a consistent string background.  The deformation
is negligible in the weak coupling region, and therefore it does
not affect the extensive properties of the theory (those
proportional to $V$). The ``Liouville wall'' \mudef\ prevents the
strings from propagating into the strong coupling region, and
provides an effective coupling constant of order $1 \over \mu$.

In the $\mu \to \infty$ limit our theory is weakly coupled and can
be analyzed using standard worldsheet methods.  The sphere
contribution is of the form $a_0\mu^2 \log \mu$, the torus is $a_1
\log \mu$ and higher corder corrections are suppressed by powers
of $1\over \mu^2$. Experience in other two-dimensional string
theories suggests that the zero point function (logarithm of the
second quantized functional integral) is of the form
 \eqn\fiform{\eqalign{
 \log \CZ =& -(a_0\mu^2 + a_1) V + f(\mu)
 +\CO({1\over V})\cr
 = &a_0\mu^2(\log\mu - V) + a_1(\log\mu - V) +
 \sum_{n=2}^\infty a_n {1\over \mu^{2n-2}} +\CO({1\over V})}}
with $f(\mu)$ a smooth function of $\mu$ which depends on the
strong coupling region of the theory.  In most of our discussion
we will be interested in the value of $a_1$.  It can be thought of
in two different ways.  First, it is the extensive part of $\log
\CZ$ which can be calculated with $\mu=0$. Alternatively, it is
the coefficient of $\log  \mu$ in $f(\mu)$.

There is no way to add such a tachyon condensate with a parameter
$\mu$ in the THO theory, and therefore the behavior of this theory
is sensitive to the physics in the strong coupling region.

\newsec{Quantization of long strings}

In this section we will study the quantization of long strings.
This quantization is similar to that of the short closed strings
but differs from it in a few crucial details.

We start by considering a single long string.  One way to quantize
it is to choose a static gauge where $\phi$ and $x$ are identified
with the worldsheet coordinates. This eliminates them as
worldsheet superfields, and therefore removes also their fermionic
super-partners $\psi_\phi$ and $\psi_x$. Equivalently, we can
follow the standard lightcone gauge fixing procedure.  Unlike
standard closed strings, the long string is not subject to
periodic boundary conditions, and therefore the zero modes of $x$
and $\phi$ are also removed.

Another difference between this quantization and the quantization
of the closed strings is that the $L_0=\bar L_0$ constraint should
not be imposed.  This constraint is associated with the periodic
boundary conditions of the closed string which are absent here.

Finally, since the string is infinite, we do not have NS and R
sectors for its left-moving fermions.  Similarly, we do not
perform a GSO projection by summing over the spin structures in
the time direction.

To summarize, the spectrum of the long string includes 24
left-moving fermions.  Clearly, a long anti-string is a long
string with the opposite orientation and its spectrum includes 24
right-moving fermions.

Note that the spectrum of the long string is the same in all our
three theories HE, HO and THO.

Now let us consider a pair of a long string and a long
anti-string.  This pair can form, as in figure 2, an infinite
folded string which ends at $\phi_0$. We would like to make
several comments about this string.

First, because of its infinite length, its energy is infinite.
With a finite cutoff $\phi > -V$ the energy of the string is
proportional to $V+\phi_0$. The $\phi_0$ dependence leads to a
force which pushes the string toward the weak coupling end,
$\phi_0 \to -\infty$. This dynamics was studied in the bosonic and
type 0 noncritical string in \MaldacenaHI. Unlike the situation
there, in our case this infinite folded string has oscillators
living on it. Below we will examine their dynamics and will see
that these oscillators can lead to important consequences.

Second, we would like to compare the infinite folded string to a
long closed string.  Locally, a very long closed string is similar
to the folded string.  The difference between them is that the
folded string is infinite, or if we regularize the space at $\phi
> -V$, its end is glued to the boundary at $V$.  This fact means
that unlike the long, but finite closed string, the folded string
is not invariant under rigid shifts of its space coordinate
$\sigma$.  Therefore, the closed string has to satisfy the
$L_0=\bar L_0$ constraint, but the long folded string does not
have to satisfy this relation.  This is the reason the fermionic
oscillators lead to physical excitations on the long folded
string, but not on the closed string.

\newsec{Anomaly cancellation with long strings}

The THO theory has gauge and gravitational anomalies.  We propose
that these anomalies can be cancelled by adding a single long
string. This anomaly cancellation can be understood in two
different ways.

The simplest way to understand it is to note that the anomalies of
the 24 spacetime right-moving fermions is cancelled by the anomaly
of the 24 fermions of the long string.  These left-moving
worldsheet fermions become left-moving spacetime fermions and
hence they cancel the anomaly.

An alternative way to understand it is the following.  We can
attempt to cancel the anomaly using the Green-Schwarz mechanism.
The Green-Schwarz mechanism involves two facts.  First, the gauge
and gravitational gauge transformations affect the $B$ field. This
is true in any theory and follows simply from the analysis of the
two-dimensional worldsheet theory.  The second element is the
Green-Schwarz term.  In two dimensions it is simply $\int B$.
Clearly, with such a term the anomalies are cancelled.  However,
with such a term the $B$ equation of motion cannot be satisfied --
there is a nonzero $B$ tadpole.  This can be fixed by adding a
long fundamental string.

More generally, if we have $W$ long strings and $\bar W$ anti-long
strings, the $B$ tadpole condition is $W=\bar W+1$ in the THO
theory ($W=\bar W-1$ in the parity transform of the THO theory)
and $W=\bar W$ in the HE and HO theories.

It is amusing to compare the effect of the Green-Schwarz term in
different dimensions.  As the number of dimensions is reduced the
term has fewer derivatives and its effect is more dramatic.  In
four dimensions it leads to the Higgs mechanism
\refs{\WittenDG,\DineXK}. Sometime it also leads to tachyonic
masses to some of the closed string modes and the need to shift
their expectation values.  In two dimensions it leads to a $B$
tadpole and the necessity to add fundamental strings.

Finally, we would like to point out that this anomaly cancellation
mechanism is similar to other examples where a space filling brane
is needed in order to cancel anomalies.  A well known example is
the IIB theory in ten dimensions where the anomaly of the theory
with an orientifold is cancelled by adding D-branes \PolchinskiTU.
Two dimensional examples which are similar to ours appeared in
four-fold compactifications of the ten dimensional type IIA theory
\SethiES.

\newsec{Compactifications, preliminaries}

In this section we will begin the discussion of the
compactifications of these theories on various twisted circles of
radius $R$.  All the compactifications of the heterotic theory are
on the same moduli space of vacua, which is parameterized by the
radius $R$ and the Wilson line around the circle \DavisQE. We are
going to limit ourselves to compactifications which preserve the
gauge symmetry. This means that the Wilson line is in the center
of the gauge group.

In all our examples the closed string vertex operators have the
form
 \eqn\veropa{\eqalign{
  &T_r=e^{-\varphi}\bar \CO_r(\half +NW) V_{N,W}\cr
  &\Psi_r = e^{-\half\varphi + i\half H }\bar \CO_r(\half +NW) V_{N,W}~,
 \qquad  p_R \ge 0  \cr
  &\tilde \Psi_r = e^{-\half\varphi - i\half H }\bar \CO_r(\half +NW)
 V_{N,W}~, \qquad p_R  \le 0\cr
 }}
with
 \eqn\nots{\eqalign{
 & V_{N,W}= e^{i {N\over R}(x+{\bar x}) + i {W\over 2} R(x-{\bar x})
  + (1-|p_R|)\phi } \cr
 &p_{R} = {N\over R} + {WR\over 2}}}
Here we have separated the coordinate $x$ to its worldsheet left
and right-moving components $\bar x$ and $ x$.\foot{We use the
same notation $x$ for the right-moving part and the full field,
hoping that this will not lead to a confusion.}  $\bar\CO_r(\bar
\Delta)$ are operators in the conjugacy class $r=0,V, S, C$ of
$Spin(24)$ in HO and THO and of $Spin(8) \times E_8$ in HE with
dimension $\bar \Delta$. Values of $(N,W)$ which do not satisfy
the inequalities of $p_R$, or lead to $\bar \Delta$ with no
$\CO(\bar \Delta)$ do not correspond to physical operators.

The different theories and the different compactifications differ
by the sets of $(r,N,W)$ in \veropa.  The values of $W$ are even
or odd integers while $N$ are either integer or half integer.
Below we will parameterize them in terms of integers as
 \eqn\nwint{N(n,w), \qquad W(n,w), \qquad {\rm with} \qquad
 n,w \in \Bbb Z }

We will be interested in the extensive part of the logarithm of
the second quantized functional integral
 \eqn\Gammadef{\log\left(\CZ(R)\right) = V\Gamma(R)+\CO(1)}
where $V$ is the volume.  Usually it is obtained in the one loop
approximation.

In all our examples we have locally
 \eqn\Gammagenf{\Gamma(R) = a R +{b \over R}}
The first term $aR$ is identified with minus the vacuum energy
density. It is not calculable in field theory, but is finite in
string theory. $a$ is independent of the details of the
compactification; i.e.\ different compactifications of the same
theory have the same the large $R$ limit of $\Gamma(R)$. Usually,
the string computation of $a$ is based on the torus amplitude with
noncompact $x$. We denote this value of $a$ by $a_{closed}$
because it is associated with closed strings. The distinction
between the true value of $a$ and the torus answer $a_{closed}$
will be clear below.

The second term $b\over R$ probes the spectrum of the theory. It
is also usually derived by a torus computation for large $R$.  We
will denote its value by $b_{closed}$.  For sufficiently large $R$
the constant $b_{closed}$ can also be calculated using a one loop
field theory calculation based on the $W=0$ modes in \veropa\
\PolchinskiZF. One way to perform this calculation is to sum
(using $\zeta$-function regularization)
 \eqn\bfiet{{b_{closed} \over R}=-\half\sum_{W=0\ {\rm terms} \atop
 {\rm in}\ \veropa} (-1)^F|p_R|}
with $F$ the spacetime fermion number.  Note that since we limit
ourselves to $W=0$, $p_R={N\over R}$.

We conclude that for sufficiently large $R$
 \eqn\Gammaclosedr{\Gamma_{world sheet}({\rm large}\
 R)=\Gamma_{closed}= a_{closed} R +{b_{closed} \over R}}

However, the careful analysis in \DavisQE\ showed that for
sufficiently small $R$ the torus amplitude $\Gamma_{world sheet}$
does not need to agree with $\Gamma_{closed}$. In fact, it was
shown that in some examples $\Gamma_{world sheet}$ exhibits a
phase transition, while $\Gamma_{closed}$ is smooth.  Furthermore,
as we will discuss below, even for large $R$ where $\Gamma_{world
sheet}=\Gamma_{closed}$ the correct value of $\Gamma$ can be
different!

The theory with compact $x$ has two shift symmetries corresponding
to the two conserved charges $N$ and $W$.  The deformation \mudef\
is the only possible deformation preserving both $N$ and $W$. In
some of our compactifications the operator \mudef\ is twisted, and
then this deformation is not present.  Other deformations break
either $N$, or $W$, or both.  For example, using $T_{r=V}$ in
\veropa\ we sometimes have the operator
 \eqn\sineliou{\mu_{sl}\bar \lambda \psi \left( V_{N=0,W=1}
 + V_{N=0,W=-1}\right) +...}
where $\mu_{sl}$ is the coupling constant, $\bar \lambda$ is one
of the left-moving fermions, $\psi$ is an appropriate linear
combination of $\psi_x$ and $\psi_\phi$ and the ellipses represent
$\CO(\mu_{sl}^2)$ corrections which are needed for $(0,1)$
worldsheet supersymmetry. This worldsheet theory is the same as
the more familiar $N=1$ super-sine-Liouville theory. Clearly,
translation invariance is preserved, but translation around the
dual circle, or $W$ conservation, is no longer a symmetry.

Another interesting deformation is by an operator constructed out
of $T_{r=0}$ in \veropa.  The simplest such deformation is
 \eqn\nsineliou{\nu \psi \left( V_{N=\half,W=-1}
 +V_{N=-\half,W=1}\right) +...}
where $\nu$ is the coupling constant.  Like the sine-Liouville
interaction \sineliou, this operator depends on $R$.  It breaks
both $N$ and $W$ conservation but leaves one linear combination of
them unbroken.  Unlike \sineliou, the operator \nsineliou\
preserves T-duality under $R \to {1\over R}$. Although we will not
analyze the detailed dynamics of the theory deformed by this
operator, we point out that its $x$ dependent part is relevant for
$\sqrt 2 -1 <R< \sqrt 2+1$, and therefore in this range we expect
its dynamics to be most interesting.  Also, at the selfdual point
$R=1$ the operator is massless -- its Liouville dressing is
similar to that of \mudef. At that point the operator \nsineliou\
is independent of the right-moving $x$ and its dependence on $\bar
x$ is through $e^{i\bar x}$, which can be thought of as a
bosonized fermion. Therefore, at this point this operator is
identical to \mudef.

\newsec{Modifying the vacuum -- changes to the worldsheet
description}

We will now examine the canonical quantization of our second
quantized theory in the $\phi$-channel, and will focus on the
asymptotic weak coupling region.  Viewing $x$ as a space
coordinate, and reducing on that circle, our system becomes a
quantum mechanical system with $\phi$ being Euclidean time.

We regularize the $\phi$ direction, $\phi \in (-V,0 )$. The
boundary conditions at the two ends $\phi=-V,0$ can be thought of
as boundary states $| s_V\rangle$ and $| s_0\rangle$.  Then, our
amplitude is the Euclidean time evolution from $\phi=-V$ to
$\phi=0$
 \eqn\eucev{\CZ= \langle  s_0| e^{-VH_\phi} | s_V\rangle}
where $H_\phi$ is the $\phi$-channel Hamiltonian. The two boundary
states $| s_V\rangle$ and $| s_0\rangle$, which appear
symmetrically in this evolution, are conceptually different. The
state $| s_0\rangle$ summarizes the behavior of the system in the
strong coupling region.  It is determined by the dynamics of the
theory. If the theory is not deformed by an interaction like
\mudef\ or others, $| s_0\rangle$ cannot be determined without a
nonperturbative definition of the theory.  However, if a term like
\mudef\ is present, the state $| s_0\rangle$ can be determined in
the $\mu \to \infty $ limit by a tree level calculation.  The
state $|s_V\rangle$ is somewhat arbitrary. It determines the
boundary conditions at the weak coupling end and through these
boundary conditions it leads to different observables.

One choice for the state $|s_V\rangle$ is the ground state. Other
choices are obtained by acting on it with creation operators.  We
identify these creation operators with the physical vertex
operators \veropa. Their energy is $|p_R|$.\foot{The branch of the
$\phi$ dressing is such that the field configuration associated
with each operator diverges at the weak coupling end $\phi \to
-\infty$ \refs{\SeibergEB,\PolchinskiMH}.  This choice of branch
corresponds in the $\phi$-channel to a creation operator.  The
opposite branch which vanishes as $\phi\to -\infty$ is an
annihilation operator.  One way to see that is to note that these
operators raise the energy of the $\phi$-channel state.  We will
see that in more detail below.} We conclude that different
correlation functions of vertex operators correspond to different
states at the weak coupling end $|s_V\rangle$, and lead to
different physical observables.

For large $R$ the low lying quanta are the $W=0$ modes in \veropa.
They are obtained by reducing the spectrum of the theory with
noncompact $x$ on a circle with appropriate boundary conditions.
In performing this dimensional reduction we should remember that
our theory is a gauge theory. The gauge fields, the graviton and
the $B$ field of the two-dimensional theory lead to three kinds of
gauge fields in the $\phi$-channel theory $\oint dx A_\phi$,
$\oint dx B_{x\phi}$ and $\oint dx G_{x\phi}$. We should now
address the effect of these gauge symmetries.

The easiest gauge symmetry to address is the one associated with
$B_{x\phi}$. It imposes conservation of the charge $W$.  There are
no closed strings with nonzero $W$ in our low energy approximation
at large $R$, so naively it has no effect. However, as we have
seen above, the anomaly of the THO theory leads to a $B$ tadpole
and the necessity to add a long string.  In our quantum mechanical
system this means that the vacuum carries nonzero $W$. Clearly,
$W=\pm 1$.  Instead of directly determining its sign, we will rely
on the consistency of the calculations below to set
$W_{vacuum}=-1$. Then the $B_{x\phi}$ gauge invariance forces us
to add a closed string vertex operator from \veropa\ with $W=1$ in
order to cancel this charge.  The addition of this vertex operator
is the $\phi$-channel way of describing the addition of a single
long string in the $x$-channel.

A more subtle effect arises from the gauge invariance of the
metric component $G_{x\phi}$.  It implements the conservation of
$N$. We will now argue that in some situations the vacuum of our
system has nonzero $N$ which should be cancelled.  The low energy
degrees of freedom in our cylinder are conformally invariant.
Therefore, it is reasonable to assign to each mode in \veropa\ the
eigenvalues $D$ and $\tilde D$ of the left-moving and right-moving
components of the target space energy momentum tensor.  We set
 \eqn\DtildeD{(D,\tilde D)=\cases{
 (p_R,0) & $p_R>0$ \cr
 (0,|p_R|) & $p_R<0$}}
The vacuum also carries such values. One way to find them is to
follow the computation of \bfiet\ separately for the left and
right-movers
 \eqn\bfieta{\eqalign{
 & E=\half\sum_{W=0\atop p_R>0} (-1)^F p_R \cr
 & \tilde E=\half\sum_{W=0\ \atop p_R<0} (-1)^F |p_R|  }}
with $\zeta$-function regularization.  Alternatively, by
identifying the boundary conditions of the spacetime fields we can
view the ground state of our system as associated with an
appropriate twist operator in the target space conformal field
theory, whose left and right dimensions are easily determined
using standard conformal field theory.  (In the target space
conformal field theory the parameters $E$ and $\tilde E$ are the
conformal dimension minus the central charge over $24$.)  The
upshot of this calculation is that we typically get $E\not=\tilde
E$, and therefore the vacuum carries momentum $N=R(E-\tilde E)$,
which should be cancelled by adding an operator with
 \eqn\addop{N_{operator}=R(\tilde E- E)}

We now discuss the effect of the gauge field $A_\phi$. The vacuum
of our system can carry $Spin(8)$ or $Spin(24)$ charges.  These
arise as follows. The target space fermions in \veropa\ with
$p_R=0$ are fermion zero modes. Quantizing them as a Clifford
algebra leads to a ground state in a representation of the gauge
group. This forces us to turn on an operator in \veropa\ in order
to form a gauge invariant state.  We will discuss this in more
detail in the specific examples below.

We conclude that in order to have an invariant state at the weak
coupling end we might need to act on the lowest energy state with
an operator from \veropa. This operator carries energy $|p_R|$.
Since our states evolve along the $\phi$ direction for ``time''
$V$, the effect of this operator insertion is to contribute to the
partition function a factor of
 \eqn\contpar{e^{-V|p_R|}}
The list of operators \veropa\ can have many different operators
with the required quantum numbers.  It is clear from \contpar\
that we should take the one with the minimum value of $|p_R|$.

We conclude that we should add another contribution to
$\Gamma_{worldsheet}$
 \eqn\addGamma{\Gamma=\Gamma_{worldsheet} -\min |p_R|}
where the minimum is among all the operators with the appropriate
charges.

What is the interpretation of this modification of the
$\phi$-channel vacuum from the point of view of the $x$-channel?
Here we calculate a trace
 \eqn\partfuop{\CZ= \Tr\ O e^{-2\pi R H_x}}
with $H_x$ the Hamiltonian for evolution along $x$ and $O$ an
appropriate operator associated with the twisted boundary
conditions.

The simplest situation is the case with nonzero $W$ insertion,
which occurs in the THO theory.  Here the Hilbert space we trace
over includes not only the closed string states, but also the long
strings and their excitations. The added operator has $p_R =
{R\over 2} + {b_{long} \over R}$ for some $b_{long}=N$, and
therefore it contributes to $\CZ$ a factor of
 \eqn\Gammalongg{\CZ_{long}=e^{V\Gamma_{long} +\CO(1)}=
 e^{-V\left({R\over 2} + {1 \over R}b_{long}\right) +\CO(1)}}
The first term ${RV\over 2}$ represents the energy of the long
string which arises from its tension.  The second term ${V\over R}
b_{long}$ represents the effect of the states in the Hilbert space
of the long string.

Our other conditions about $N$ and the gauge representation can be
interpreted as follows\foot{We thank J.~Maldacena for a useful
discussion about this point.}.  If we want to view our theory as a
free theory in the bulk of space we need to ensure proper boundary
conditions at the strong and weak coupling ends of space.  We
impose that no charge or energy emerges from or leaves the system
at the strong coupling end.  With Lorentzian signature this means
that all scattering processes preserve energy and charge; i.e.\
the total energy flux and charge which is injected into the system
in the past must come back in the future \MaldacenaHE.  (This
requirement is satisfied with any local boundary conditions at
infinity without more degrees of freedom there, but is weaker than
such locality.)  This means that the integrated energy flux and
charge current over all time must vanish
 \eqn\intfl{\int dx\ T_{x\phi} = \int dx\ J_\phi =0}

The naive calculation of the trace and the worldsheet results
assume that the left and right-moving modes are independent.
Instead, the exact eigenstates of the Hamiltonian are linear
combinations of incoming and outgoing modes such that conditions
\intfl\ are satisfied. The precise linear combination of these
modes determines the S-matrix of the theory. In the Euclidean
theory with compact $x$ the conditions \intfl\ are the same as the
invariance conditions of the state that we used above.

Finally, we would like to explain why it is always possible to fix
the invariance of the naive lowest energy state with a vertex
operator. This follows from the fact that all these
compactifications are connected \DavisQE.  We can start with a
simple circle compactification and continuously (perhaps crossing
phase transitions) move to other compactifications.  By doing that
an invariant ground state undergoes spectral flow to another
invariant state.  This state can have higher energy than the naive
ground state.  In that case it is obtained from it by acting with
a creation operator.

\newsec{Phase transitions associated with massless fermions}

Consider a situation where as we vary $R$, at some point a massive
complex fermion becomes massless and then becomes massive again.
The system includes a complex fermion operator $\Psi$ with the
Lagrangian and Hamiltonian
 \eqn\laghamfer{\eqalign{
 \CL_\phi=& i \Psi^\dagger \partial_\phi \Psi + m(R) \Psi^\dagger
 \Psi\cr
 H_\phi=& {m(R)\over 2}\left( \Psi^\dagger \Psi - \Psi
 \Psi^\dagger \right)
 }}
The subscript $\phi$ denotes the fact that these are the
Lagrangian and Hamiltonian in the $\phi$-channel.  In the examples
below the fermion mass is
 \eqn\mofrf{m(R)={R\over 2} - {1\over 2 R}}
The quantization of such a complex fermion leads to two states
$|\pm\rangle$ with energy $\pm \half m(R) $. For $m>0$ ($R>1$) the
lowest energy state is $|-\rangle$ and for $m<0$ ($R<1$) it is
$|+\rangle$. Therefore the ground state energy is $- \half
|m(R)|$.  As a check, we will see in the examples below that for
$R>1$ only the operator $\Psi^\dagger$ is present in the list of
vertex operators \veropa, while for $R< 1$ only the operator
$\Psi$ is associated with a vertex operator.  We interpret this to
mean that the other operator annihilates the ground state, and
therefore it is not included in the list \veropa.

The worldsheet calculations are consistent with this
interpretation of the ground state energy.  The result, which
includes the contribution of all the modes, is
 \eqn\fermionphaset{\Gamma_{worldsheet}= \half |m(R)| + f(R)}
with $f(R)$ a smooth function.

This is the correct answer, if we ignore the various gauge
symmetries in the problem.  We have already seen that possible
charges of the vacuum can modify this expression for $R>1$. In all
our examples the complex fermion carries nonzero $N$ and $W$, and
therefore the Lagrangian \laghamfer\ should also include the
coupling to the appropriate gauge fields. Therefore, the states
$|\pm\rangle$ have opposite charges and hence the lowest energy
state for $m>0$ ($R>1$) and for $m<0$ ($R<1$) have different
charges. Consequently, gauge invariance forces us to act on at
least one of them, and possibly both, with some vertex operator.

The simplest example is when for $R>1$ we do not need to act with
any operator because the vacuum is invariant. Then, for $R<1$ we
must act with the fermion operator $\Psi$ in order to change the
state. This raises the energy of the state by $|m(R)|$ and smooths
the transition:
 \eqn\fermionphaseta{\eqalign{
 \Gamma=&\cases{
 \Gamma_{worldsheet} & $R>1$ \cr
 \Gamma_{worldsheet}-|m| & $R<1$
 }\cr
 =&\half m(R) + f(R)
 }}

Let us discuss this phenomenon in the $x$-channel and focus on the
region where $m(R)$ is small.  More precisely, we study the limit
 \eqn\mVlim{m \to 0\ , \qquad V\to \infty \ , \qquad mV={\rm
 fixed}}
In this limit we can focus on the two states $|\pm \rangle $ of
the $\phi$-channel picture. For $m>0$ the closed string partition
function $\CZ_{closed}$ gives a good approximation to the full
answer.  It is corrected by the contribution of the long string
states. The lowest of them is associated with the fermion
$\Psi^\dagger$. Therefore we expect the partition function to be
of the form
 \eqn\fermparx{\CZ_{worldsheet}=\CZ_{closed}\left(1 + c(m)
 e^{-mV }+ ...\right)}
where the term $c(m) e^{-Vm}$ represents the contribution of a
long string in the system.  The exponential factor $ e^{-Vm}$
includes the extensive part of the logarithm of the partition
function and the prefactor $c(m)$ is a finite size effect.  Since
this string is fermionic, there is no contribution of the form
$e^{-nVm}$ which could have originated from $n>1$ long strings.
The ellipses in \fermparx\ represent possible contributions which
are negligible in the limit \mVlim\ including the effect of other
states which are suppressed by $e^{-V h}$ for some constant $h\gg
|m|$.

Now, let us examine the limit $|mV|\to \infty$ . For $m>0$ the
long string contribution vanishes. But for $m<0$ the second term
dominates and $\Gamma_{worldsheet}(R<1) = \Gamma_{closed} - m(R)$
as in \fermionphaset.

However, this neglects the constraints from gauge invariance. In
the simplest situation, where the $\phi$-channel vacuum is gauge
invariant the second term in \fermparx\ should not be included and
we have $\CZ=\CZ_{closed}$ for all $R$. Alternatively, the correct
contribution for $R<1$ is the worldsheet answer
$\Gamma_{worldsheet}$ plus the contribution of a long string
\fermionphaseta. This situation is similar to the more familiar
examples including the D0-D8 system
\refs{\PolchinskiSM\HananyIE\BanksZS-\BachasKN}, where a creation
of a fundamental string makes the vacuum energy constant when a
D0-brane passes through a D8-brane.

Below we will encounter more complicated situations, where the
gauge charges force us to act with an operator both above and
below the transition point in such a way that the transition is
not smoothed out.

\newsec{Phase transitions associated with massless bosons}

In this section we will consider a situation where as $R$ is
varied a massive complex scalar becomes massless and then massive
again.

\subsec{$\phi$-channel interpretation}

In this case the nonanalytic part of $\Gamma$ is determined by the
dynamics of a light complex scalar field $\Phi$.  Its effective
one-dimensional Lagrangian describes a complex harmonic oscillator
 \eqn\effth{\CL_\phi=  |\partial_\phi \Phi|^2 +  m(R)^2| \Phi|^2}
where in our examples
 \eqn\mofRi{m(R)={R\over 2} -{1\over 2R}}
The boundary conditions at $\phi=-V,0$ are summarized by
(boundary) states $|s_{V,0}\rangle$ in the $\phi$-channel Hilbert
space. In this language the functional integral is easily
evaluated as a sum over intermediate states. Gauge invariance
forces us to consider boundary states which are invariant under
the $U(1)$ gauge symmetry $\Phi \to e^{i\alpha }\Phi$.  We assume,
for simplicity, that our vacuum is $U(1)$ invariant (more
complicated situations will be discussed below). Then, only the
$U(1)$ invariant states $|n\rangle$ with energy $(2 n+1) |m|$
contribute
 \eqn\genZ{\eqalign{
 \CZ_{oscillator}= &\langle s_0|e^{-VH_\phi}|s_V\rangle =
 \sum_{n=0}^\infty c_n(m)  e^{-(2 n+1) |m|V}
 \cr
 c_n(m)=& \langle s_0| n \rangle \langle n | s_V\rangle
 }}
where $H_\phi$ is the $\phi$-channel Hamiltonian of the Lagrangian
\effth. The $V$ dependence of each term $ e^{- (2 n+1) |m|V}$ is
independent of the boundary conditions and depends only on the
energy of the state $|n\rangle $. However, the prefactors $c_n$
are ``finite volume corrections'' which depend on the boundary
conditions.

Here are three simple illustrative examples with $|
s_{V,0}\rangle$ being $\Phi$ eigenstates with zero eigenvalues or
$P_\Phi$ eigenstates with zero eigenvalue
 \eqn\propsumc{\eqalign{
 \langle \Phi=0|e^{-VH_\phi}|\Phi=0\rangle = &{m \over 2\pi
 \sinh (m V)}= {|m| \over \pi } \sum_{n=0}^\infty  e^{-(2 n+1)
 |m|V} \cr
 \langle P_\Phi=0|e^{-VH_\phi}|P_\Phi=0\rangle=&{1 \over 2\pi m
 \sinh (m V)}= { 1\over \pi  |m|} \sum_{n=0}^\infty  e^{-(2 n+1)
 |m|V} \cr
 \langle P_\Phi=0|e^{-VH_\phi}|\Phi=0\rangle =& {1 \over 2\pi
 \cosh (m V)}= { 1\over \pi  } \sum_{n=0}^\infty (-1)^n
 e^{-(2 n+1) |m|V} }}

As we will see in the examples below, our string problem has
T-duality symmetry which maps $m\to -m$. Furthermore, in the HE
and HO theories at the selfdual point $m=0$ the T-duality
transformation is part of an enhanced non-Abelian target space
symmetry (gauge symmetry of the $\phi$-channel problem).
Therefore, it must be an exact symmetry of the system \DineVU, and
hence the states $|s_{V,0}\rangle$ should be independent of the
sign of $m$.  Hence, the coefficients $c_n(m)$ and the  partition
function \genZ\ should also be independent of the sign of $m$.
This is the case in the examples \propsumc.

If $c_0(m)\not= 0$, the large $V$ limit of $\CZ_{oscillator}$ is
dominated by the ground state
 \eqn\largeVg{\lim_{V\to \infty}\CZ_{oscillator} = c_0(m) e^{- |m|V}}
It is nonanalytic in $m$ around $m=0$.  However, the finite $V$
expressions are analytic in $m$ around $m=0$.  Using \genZ\ we
find for normalizable states
 \eqn\smallma{\lim_{m \to 0}\CZ_{oscillator} = \sum_{n=0}^\infty
 c_n(m)= \langle s_0| s_V\rangle }
which is finite.  Even for delta-function normalizable states, as
in the examples \propsumc, it is easy to check that $\lim_{m \to
0}\CZ_{oscillator}$ is either finite or has a double pole.
Therefore, for finite $V$ we can analytically continue the
expressions from positive to negative $m$.

Now we embed this harmonic oscillator in our string problem.  We
assume that both $| s_0\rangle$ and $| s_V\rangle$ have nonzero
overlap with the harmonic oscillator vacuum $| 0\rangle$ so that
$c_0$ in \genZ\ is nonzero.  Including all the other modes we
conclude that the infinite $V$ worldsheet expression is
nonanalytic . It is given by
 \eqn\bosonnphaset{\Gamma_{worldsheet}= -|m(R)| + f(R)}
with $f(R)$ a smooth function.  This will turn out to be
consistent with explicit worldsheet calculations and T-duality.

\subsec{$x$-channel interpretation -- puzzling thermodynamics}

So far we discussed the transition from the point of view of the
$\phi$-channel.  We now discuss it in the $x$-channel, where
$\phi$ is viewed as space and $x$ as time.  As we will see in the
examples below, this transition happens whenever our system is
compactified on a thermal circle; i.e.\ twisted by $(-1)^F$ with
$F$ the spacetime fermion number.  Usually this means that the
$x$-channel system is at finite temperature with the partition
function
 \eqn\thermpax{\CZ = \Tr e^{-2\pi R H_x}}
where $H_x$ is the $x$-channel Hamiltonian.

The expression \thermpax\ has two immediate standard consequences.
First, the partition function can be written as a sum of positive
numbers, and therefore $\CZ>0$.  Second, the expectation value of
positive quantities like $\left\langle \left(E-\langle
E\rangle\right)^2 \right\rangle$ should be positive, and therefore
the specific heat must be positive
 \eqn\standthe{c \sim R^2 \partial_R^2 \Gamma  >0}

For simplicity we focus on situations where the $\phi$-channel
ground state is invariant and no long strings are needed. This
will be the case in the HE and HO theories.  The extension to more
complicated theories like THO is straightforward.

We start the discussion for $R>1$ where $m>0$. Here
$\Gamma_{closed}$ which is computed as $\Tr e^{-2\pi R H_x}$ with
the closed string spectrum agrees with $\Gamma_{worldsheet} $.
However, an important subtlety has to be stressed. In addition to
the closed strings, the Hilbert space includes also the long
folded strings of figure 2. As we said above, such states have
large energy proportional to the large volume $V$, and therefore
for sufficiently large $R$ (small temperature) they do not
contribute to the large $V$ limit of the partition function.
Because of the degrees of freedom on them they also have large
entropy proportional to $V\over R^2$. This has no effect for
$R>1$. However, for $R=1$ the gas of such long strings can
contribute to the partition function. This explains why the answer
$\Gamma_{closed}$ is correct for $R>1$ but is wrong for $R\le 1$.
It simply misses the contribution of these states in the Hilbert
space.

Let us analyze these folded long strings in more detail.  As we
said above, the energy of such a folded string is proportional to
the length $V+\phi_0$ and diverges in the thermodynamic limit
$V\to \infty$. This dependence on $\phi_0$ leads to an attractive
force proportional to $R$ toward the weak coupling end.  Hence
such a folded string is unstable and retracts back to the cutoff
at the weak coupling end \MaldacenaHI. However, unlike the
examples studied in \MaldacenaHI, in our case, such a folded
string has oscillators living on it, and they lead to entropy.
This produces a force proportional to $1\over R$ pushing $\phi_0
\to +\infty$. For $R>1$ the attraction to $\phi_0 \to -\infty$ is
the dominant effect, but for $R=1$ these two forces are exactly
balanced.  It seems that for $R<1$ these strings are attracted to
the strong coupling end and our system will have a condensate of
such long strings.

This long string condensation picture is similar to the picture of
the Hagedorn transition in higher dimensions.  There the
transition is associated with the condensation of long closed
strings \refs{\AtickSI,\HorowitzJC}.  In our case, the closed
strings do not have the necessary entropy to lead to Hagedorn
behavior. Instead, a closed string which is stretched all the way
to infinity is the folded string.  It has the necessary entropy to
lead to a transition.

We would like to study the region near the transition point at
large but finite $V$.  Therefore, we study the limit \mVlim
 \eqn\mVlima{m \to 0\ , \qquad V\to \infty \ , \qquad mV={\rm
 fixed}}
Our discussion of the $\phi$-channel picture suggests that for
finite $V$ the $m$ dependence can be analytically continued to
$m<0$ ($R<1$).  Before we see where this analytic continuation
leads us, we examine the partition function for $m>0$ ($R>1$) in
the limit \mVlima, making the intuitive picture above more
quantitative. The $\phi$-channel expression \genZ\ shows that for
$m>0$ the partition function is
 \eqn\bosonZRb{\CZ\approx \CZ_{closed}\left (1+ \sum_{n=1}^\infty
 {c_n\over c_0} e^{-2n mV}\right) }
where we have identified the first term as $\CZ_{closed}$.  In the
$\phi$-channel picture the various terms in the sum in \bosonZRb\
are associated with the excited states of the harmonic oscillator.
We now interpret them in the $x$-channel as representing the
contributions of different number of pairs of long strings. Each
pair has a partition function
 \eqn\partpair{\CZ_{pair}\sim e^{-2m V}}
where the factor of $2$ is because of the string and the
anti-string.  As we have already remarked, for $m(R)={R\over 2}-{1
\over 2R}$ the first term $ {R\over 2} $ is due to the energy of
the long string and the second term $1\over 2R$ represents its
entropy.

In order for this $x$-channel interpretation to be sensible, we
need $c_n\over c_0$ to be positive real numbers for all $n$.
Clearly, the generic boundary state in \genZ\ does not satisfy
this requirement (see, e.g.\ the examples \propsumc), but boundary
conditions at $\phi=-V,0$ which are local in $x$ satisfy it. Let
us examine whether such local boundary conditions are possible.

We have discussed the expansion \bosonZRb\ for $m>0$.  Since we
are studying the finite volume system, it is reasonable that it
converges for most $m$ and defines an analytic function of $m$.
This expectation is supported by the $\phi$-channel expression
\genZ\ which defines such an analytic function (see e.g.\ the
examples \propsumc).  Furthermore, as we mentioned above, our
system should be invariant under the T-duality transformation
which acts as $m \to -m$. This leads to \largeVg\ and
\bosonnphaset
 \eqn\totalgamma{\Gamma = \lim_{V\to \infty} {1\over V} \log (\CZ) =
 \cases{
 \Gamma_{closed} & $m>0$ \cr
 \Gamma_{closed}- 2 |m| & $m<0$ \cr
 }}

However, \totalgamma\ exhibit a problem -- the system has a first
order phase transition at $m=0$ ($R=1$) whose latent heat is
negative.  It is easy to see that any smooth finite $V$ expression
which asymptotes to \totalgamma\ as $V\to \infty$ violates
\standthe; i.e.\ it has negative specific heat.  Since \standthe\
follows from the positivity of the weights in the trace over the
Hilbert space, our answer \totalgamma\ cannot arise in any
thermodynamical system.

We conclude that for infinite $V$ and negative $m$ there is a
conflict between the $\phi$-channel and the $x$-channel pictures.
For finite $V$ this conflict arises for all $m$ but for positive
$m$ the problem is exponentially small in $V$. This conflict can
be traced back to the contribution of the ground state energy of
the harmonic oscillator in the $\phi$-channel being $|m|$ and the
long string contribution $\Gamma_{long} = -|m|=-\left|{R\over 2}
-{1\over 2R}\right| $ and $\Gamma_{pair}=2\Gamma_{long}= -\left|R
-{1\over R}\right|$  all with absolute values. On the other hand,
the $x$-channel picture has, as in \partpair\ $\Gamma_{pair}=-2m =
-\left(R -{1\over R}\right)$ without absolute value (recall the
interpretation of the first term $R$ as the associated with the
energy and the second term ${1\over R}$ as associated with the
entropy).

We suggest that the boundary states $|s_{V,0}\rangle $ which are
invariant under $m \to -m$ and in particular the ground state
$|0\rangle$ are associated with boundary conditions which are
nonlocal in $x$. Such nonlocal boundary conditions do not satisfy
the thermodynamical positivity of the $x$-channel picture. A
simple example is the third inner product in \propsumc.  The
expansion in the right hand side has both positive and negative
quantities and cannot be given an interpretation as $\Tr e^{-2\pi
R H_x}$ in the $x$-channel.

We conclude that for negative $m$ the $\phi$-channel quantization
is nonlocal in $x$ and cannot be interpreted as thermodynamics in
the $x$-channel.  Instead, it has another interpretation with
$\phi$ as space describing the thermodynamics of the T-dual
system, whose radius is $1\over R$ and is associated with local
evolution along the T-dual circle.  For $R<1$ the expressions
$\Gamma_{long} = -|m|=-\left|{R\over 2} -{1\over 2R}\right| $ and
$\Gamma_{pair}=2\Gamma_{long}= -\left|R -{1\over R}\right|$
describe not a long string and a long pair but their T-dual -- a
long string and a long pair in the T-dual theory.

To summarize, the $\phi$-channel picture leads to a sensible
expression for the partition function of the Euclidean circle
problem for all $R$.  This expression agrees with thermodynamics
in the $x$ channel only for $R>1$.  For $R<1$ it agrees with
thermodynamics of the T-dual theory on a circle of radius $1\over
R$.  However, these two thermodynamics interpretations cannot be
continued to higher temperatures.  We suggest that the sensible
answers derived from the $\phi$-channel are associated with
nonlocal boundary conditions in $x$ and this nonlocality prevents
us from having thermodynamics at higher temperatures.  We will
return to this point in the conclusions

\newsec{Untwisted circle}

We now turn to various examples making our general discussion
above more concrete.

The simplest compactification without a twist of HO and HE  has
the vertex operators \DavisQE
 \eqn\twer{\eqalign{
  &T_V=e^{-\varphi}\bar \CO_V(\half +nw) V_{n,w}\cr
  &\Psi_S = e^{-\half\varphi + i\half H }\bar \CO_S(\half +nw) V_{n,w}~,
 \qquad  p_R \ge 0  \cr
  &\tilde \Psi_C = e^{-\half\varphi - i\half H }\bar \CO_C(\half +nw)
 V_{n,w}~, \qquad p_R  \le 0\cr
  }}
In the THO theory we have \Davisup
 \eqn\thocirc{\eqalign{
  &T_C=e^{-\varphi}\bar \CO_C(\half +nw) V_{n,w}\cr
  &\Psi_S = e^{-\half\varphi + i\half H }\bar \CO_S(\half +nw)
  V_{n,w}~,\qquad  p_R \ge 0  \cr
  &\tilde \Psi_V = e^{-\half\varphi - i\half H }\bar \CO_V(\half
  +nw)V_{n,w}~, \qquad p_R  \le 0\cr
  }}
These three compactifications are invariant under the T-duality
transformation $R \to {2 \over R}$, with enhanced $SU(2)$ symmetry
at the selfdual point $R=\sqrt 2$.

The field theory calculation of $\Gamma_{closed} $, based on the
spectrum of the theory in noncompact spacetime is straightforward
\refs{\DavisQE,\Davisup}
 \eqn\gammac{ \Gamma_{closed} =\cases{
  R+{2\over R} & HO \cr
 0 & HE \cr
  -{R \over 2}-{1\over R}  & THO }}
The analysis of \DavisQE\ shows that in these theories the
worldsheet calculation leads to the same answer for all $R$
 \eqn\gammacwsc{\Gamma_{worldsheet}(R)=\Gamma_{closed}(R)}

Let us examine whether the $\phi$-channel ground state has to be
modified. First, in HO and HE we must have $W=0$, and in THO we
should add $W=1$. Second, it is easy to check using \bfieta\addop\
that in HO and HE we need $N=0$, while in THO we need $N=1$.

Next we check the gauge charges of the vacuum. In the HO theory
there are no fermion zero modes, and therefore there is no
subtlety in the quantization.

The HE theory has two fermion zero modes $\Psi_S(N=W=0)$, whose
quantization leads to ${\bf 8}_V \oplus {\bf 8}_C$, and $\tilde
\Psi_C(N=W=0)$, whose quantization leads to ${\bf 8}_V \oplus {\bf
8}_S$.  There is a unique gauge invariant state in ${\bf 8}_V
\otimes {\bf 8}_V$, and therefore no modification of the spectrum
is needed.

The THO theory has one fermion zero mode $\tilde \Psi_V(N=W=0)$
whose quantization leads to ${\bf 2^{11}}_S \oplus {\bf
2^{11}}_C$.  The lack of gauge invariance of this representation
is cancelled by acting on the lowest energy state with a vertex
operator.  Combining this information with the fact that the
operator should have $N=W=1$, we find that this operator is
$T_C(N=W=1)$ or $\Psi_S(N=W=1)$. We interpret these operators as
associated with the long string that must be added to the system.

We conclude that
 \eqn\gammact{ \Gamma =\cases{
  R+{2\over R} & HO \cr
 0 & HE \cr
  -{R \over 2}-{1\over R} -\left({R \over 2}+{1\over R} \right)
  = -R -{2\over R} & THO }
  }
As a check, note that this answer is invariant under the T-duality
transformation $R \to {2 \over R}$.

Since as we vary $R$ no mode becomes massless and correspondingly
$\Gamma_{long}$ is always positive, there cannot be a phase
transition in this system, and therefore the expressions \gammact\
are correct for all $R$.

\newsec{Twisted circles}

\subsec{Twist by $(-1)^{f_L}$}

We now consider compactifications which are twisted by
$(-1)^{f_L}$ with $f_L$ the worldsheet fermion number of the
left-movers.  Such twists are possible only if the tachyon
deformation \mudef\ vanishes. Also, this twist breaks the
spacetime parity symmetry. One way to see that is to note that
$(-1)^{f_L}$ acts differently on the representations with $C$ and
$S$, which are exchanged by spacetime parity.

The spectrum of vertex operators in HE and HO is \DavisQE
 \eqn\newtwist{\eqalign{
 &T_V = e^{-\varphi} \bar{\cal O}_V(\half+(2n+1)w)
 V_{n+{1\over 2},2w} \cr
 & T_C = e^{-\varphi} \bar{\cal O}_C (\half + n(2w+1)) V_{n,2w+1}
 \cr
 &\Psi_S = e^{-\varphi/2 + iH/2} \bar{\cal O}_S (\half + 2nw)
 V_{n,2w}~ , \quad\quad\quad\quad\quad\quad~~~~~~ p_R\geq 0~~\cr
 &\Psi_0 =e^{-\varphi/2 + iH/2} \bar{\cal O}_0 (\half +
 (n+\half)(2w+1)) V_{n+{1\over 2},2w+1}~,~~p_R\geq 0~~\cr
 &\tilde\Psi_C =e^{-\varphi/2 - iH/2} \bar{\cal O}_C (\half +
 (2n+1)w) V_{n+{1\over 2},2w}~,\quad\quad\quad ~~~ p_R\leq 0\cr
 &\tilde\Psi_V =e^{-\varphi/2 - iH/2} \bar{\cal O}_V (\half +
 n(2w+1))V_{n,2w+1}~,\quad\quad\quad ~~~p_R\leq 0 }}
and in THO it is \Davisup
 \eqn\thospin{\eqalign{
 &T_V = e^{-\varphi} \bar{\cal O}_V(\half+n(2w+1)) V_{n,2w+1} \cr
 & T_C = e^{-\varphi} \bar{\cal O}_C (\half + (2n+1)w)
 V_{n+\half,2w} \cr
 &\Psi_S = e^{-\varphi/2 + iH/2} \bar{\cal O}_S (\half + 2nw)
 V_{n,2w}~, \quad\quad\quad\quad\quad\quad~~~~~~ p_R\geq 0~~\cr
 &\Psi_0 =e^{-\varphi/2 + iH/2} \bar{\cal O}_0 (\half +
 (n+\half)(2w+1)) V_{n+{1\over 2},2w+1}~,~~~p_R\geq 0~~\cr
 &\tilde\Psi_C =e^{-\varphi/2 - iH/2} \bar{\cal O}_C (\half +
 n(2w+1)) V_{n,2w+1}~,\quad\quad\quad ~~~~ p_R\leq 0\cr
 &\tilde\Psi_V =e^{-\varphi/2 - iH/2} \bar{\cal O}_V (\half +
 (2n+1)w)V_{n+\half,2w}~,\quad\quad\quad ~~~p_R\leq 0 }}
HE is selfdual (with $V\leftrightarrow C$) under $R \to {1\over
R}$ with an enhanced $Spin(10) \times E_8$ symmetry at the
selfdual point $R=1$. The HO theory compactified on a circle of
radius $R$ is dual to the THO on a circle of radius $1\over R$. At
the point $R=1$ these theories include the worldsheet current
$e^{-\varphi/2 + iH/2} V_{\half,1}=e^{-\varphi/2 + iH/2 + ix}$,
which leads to an anticommuting target space (super)symmetry. At
this point these theories are special cases of the noncritical
superstring construction of \KutasovUA.

The field theory calculation of the closed strings leads to
\refs{\DavisQE,\Davisup}
 \eqn\gammatwc{
 \Gamma_{ closed}  =  \cases{
 -{1\over 2 R}  & HE \cr
 R - {1\over R}& HO \cr
 -{R\over 2}+{1\over 2 R} & THO }}
and the worldsheet answers are \refs{\DavisQE,\Davisup}
  \eqn\gammatwry{\eqalign{
  \Gamma_{ worldsheet}(R>1) =& \Gamma_{ closed}(R)\cr
   \Gamma_{ worldsheet}(R<1) =& \Gamma_{ closed}(R)
 -\left({R\over 2}-{1\over 2R}\right) =  \cases{
 -{R\over 2}  & HE \cr
  {R\over 2} - {1\over 2R}& HO \cr
 -R + {1\over  R} & THO }}}
where we see the effect of the massless fermion at $R=1$.

We now consider the necessary modifications due to gauge
invariance, and start with the situation for $R>1$. There is no
need to change the HO theory. The THO theory needs an insertion of
a long string ($W=1$) with $N=-\half$; i.e.\ we act on the lowest
energy state with $\Psi_0(N=-\half, W=1)$.

The HE theory is more interesting.  Here we need to add an
operator with $W=0$ and $N=-\half$.  This is our only example
where the added operator has $W=0$, and therefore the modification
of the theory cannot be interpreted as adding a long string.  For
generic $R$ we have the fermion zero mode $\Psi_S(N=W=0)$, whose
quantization leads to a ground state in ${\bf 8}_V \oplus {\bf
8}_C$. Therefore, we have to act on the $\phi$-channel lowest
energy state with $T_V(N=- \half, W=0)$ or $\tilde
\Psi_C(N=-\half, W=0)$. The fact that we can achieve gauge
invariance with zero $W$ is consistent with the fact that the
noncompact theory does not have long strings.

We conclude that for $R>1$
 \eqn\gammatw{\Gamma(R>1)  = \cases{
 -{1\over 2 R} -{1\over 2R} =-{1\over R} & HE \cr
 R - {1\over R}& HO \cr
 -{R\over 2}+{1\over 2 R}-\left({R\over 2} - {1\over 2R}\right)=
 -R+{1\over  R} & THO }}

Let us now move to $R<1$.  In all these theories a complex fermion
becomes massless at $R=1$.  Therefore, as we explained above, the
change in the quantum numbers of the lowest energy state forces us
to act with $\Psi_0(N=\half, W=-1)$. In the HO theory this adds a
long string and in THO this removes the added long string. In the
HE theory we act on the naive ground state with the lowest
dimension operator with the quantum numbers of $\Psi_0(N=\half,
W=-1)\ T_V(N=- \half, W=0) $ which is $ \tilde \Psi_V(N=0, W=-1)$,
or with the lowest dimension operator with the quantum numbers of
$\Psi_0(N=\half, W=-1)\ \tilde\Psi_C(N=- \half, W=0) $ which is
$T_C(N=0, W=-1)$.  This adds an anti-long string ($W=-1$), but
since it has $N=0$, its T-dual version does not involve a long
string.

We conclude that
 \eqn\gammatwr{ \Gamma(R<1)= \Gamma_{ worldsheet}(R<1)- \cases{
 {R\over 2}  & HE \cr
 -{R\over 2} + {1\over 2 R}& HO \cr
 0 & THO }
 =  \cases{
 -R & HE \cr
 R - {1\over R}& HO \cr
 -R+{1\over  R} & THO }}
Comparing with \gammatw\ we see that HE is indeed selfdual and HO
and THO are T-dual to each other under $R\to {1\over R}$.
Evidently, the phase transition in HO and THO at $R=1$ is smoothed
out, but the HE theory still has a phase transition there.

\subsec{Twist by $(-1)^{f_L+F}$}

Another possibility is to twist the circle by $(-1)^{f_L+F}$,
where $F$ is the spacetime fermion number. In the HO and HE theory
this twisted compactification is the same as the one with
$(-1)^{f_L}$ (section 10.1). The vertex operators of THO are
\Davisup
 \eqn\thovect{\eqalign{
 &T_C=e^{-\varphi}\bar \CO_C(\half +(2n+1)w) V_{n+\half,2w}  \cr
 &T_S=e^{-\varphi}\bar \CO_S(\half +n(2w+1)) V_{n,2w+1} \cr
 &\Psi_S = e^{-\half\varphi + i\half H }\bar \CO_S(
  \half +(2n+1)w) V_{n+\half,2w}~,\qquad ~~~~~~~p_R \geq 0  \cr
 &\Psi_C = e^{-\half\varphi + i\half H }\bar \CO_C(
 \half +n(2w+1)) V_{n,2w+1}~,\qquad ~~~~~~~p_R \geq 0  \cr
 &\tilde \Psi_0 = e^{-\half\varphi - i\half H }\bar \CO_0(
 \half +(n+\half)(2w+1)) V_{n+\half,2w+1}~, \quad p_R \leq 0  \cr
 &\tilde \Psi_V = e^{-\half\varphi - i\half H } \bar \CO_V(\half
 +2nw)  V_{n,2w}~, \qquad \qquad ~~~~~~~~~~~~p_R\leq 0
  }}
This compactification is selfdual (with $S \leftrightarrow C$)
under $R \to {1\over R}$ with enhanced $Spin(26)$ symmetry at the
selfdual point $R=1$.  At this point the theory includes the
worldsheet current $e^{-\varphi/2 -iH/2}
V_{-\half,-1}=e^{-\varphi/2 - iH/2 - ix}$, which leads to an
anticommuting target space (super)symmetry. Again, this is a
special cases of the noncritical superstring construction of
\KutasovUA.

Here \Davisup
 \eqn\gammatwag{\eqalign{
 \Gamma_{closed}  =& -{R\over 2}-{1\over  R}\cr
 \Gamma_{worldsheet}(R>1) =& \Gamma_{closed} (R>1)\cr
 \Gamma_{worldsheet}(R<1) =&
 \Gamma_{closed} (R<1) -\left({R\over 2}-{1\over 2R}\right)
  =-R -{1\over 2R}}}
where again we see the effect of the massless fermion at $R=1$.

For $R>1$ we have to act on the lowest energy state with an
operator with $W=1$ and $N=1$. The $Spin(24)$ quantum numbers are
determined by noting that $\tilde\Psi_V(N=W=0)$ is the only
fermion zero mode. Its quantization leads to ${\bf 2^{11}}_S
\oplus {\bf 2^{11}}_C$. Therefore, we must act with $T_S(N= W= 1)$
or $\Psi_C(N=W=1)$ and
 \eqn\gammatwaaa{ \Gamma(R>1)  = -{R\over 2}-{1\over  R} -\left(
 {R\over 2}+{1\over  R}\right) = -R-{2\over  R} }

For $R<1$ we must also act with $\tilde \Psi_0(N=-\half, W=1)$.
Therefore, we act on the naive ground state with $\tilde
\Psi_0(N=-\half, W=1)\ T_S(N= W= 1) \sim \Psi_S(N=\half, W=2) $ or
$\tilde \Psi_0(N=-\half, W=1)\ \Psi_C(N=W=1) \sim T_C(N=\half,
W=2)$.  This can be interpreted as having two long strings in the
target space. The fact that $N=\half$ shows that in the T-dual
picture there is only one long string. Hence
 \eqn\gammatwarr{ \Gamma(R<1)  = \Gamma_{worldsheet}(R<1)-
 \left({1\over 2R} + R\right)= -2R-{1\over R} }
Comparing with \gammatwaaa\ we see that the final answer is
selfdual.

\newsec{Thermal circle}

Here we compactify our theories on a thermal circle; i.e.\ we
twist by $(-1)^F$ with $F$ the spacetime fermion number.

The spectrum of operators of the HO and HE theories is \DavisQE\
 \eqn\twert{\eqalign{
  &T_V=e^{-\varphi}\bar \CO_V(\half +2nw) V_{n,2w}
  \cr
  &T_0=e^{-\varphi}\bar \CO_0(\half +(n+\half)(2w+1)) V_{n+\half,2w+1}
  \cr
  &\Psi_S = e^{-\half\varphi + i\half H }\bar \CO_S(
  \half +(2n+1)w) V_{n+\half,2w}~,\qquad ~~p_R \geq 0  \cr
  &\Psi_C = e^{-\half\varphi + i\half H }\bar \CO_C(
  \half +n(2w+1)) V_{n,2w+1}~,\qquad ~~p_R \geq 0  \cr
  &\tilde \Psi_S = e^{-\half\varphi - i\half H }\bar \CO_S(
  \half +n(2w+1)) V_{n,2w+1}~, \qquad ~~p_R \leq 0  \cr
  &\tilde \Psi_C = e^{-\half\varphi - i\half H }
  \bar \CO_C(\half +(2n+1)w) V_{n+\half,2w}~, \qquad ~p_R \leq 0 \cr
  }}
These compactifications are T-duality invariant under $R \to
{1\over R}$. This duality is guaranteed by an enhanced symmetry at
the selfdual point $Spin(24) \times U(1) \to Spin(26)$ and
$Spin(8) \times E_8 \times U(1)\to Spin(10) \times E_8$.

It is interesting to compare this compactification with the
untwisted circle compactification \twer\ of the HO theory.  The
closed string spectrum of HO in $\Bbb R^2$ includes only bosons,
and therefore it is not affected by the thermal boundary
conditions. Nevertheless, the thermal circle \twert\ differs from
the untwisted circle \twer.  One way to understand it is to note
that the two different twists around the circle act differently on
the long strings. This difference manifests itself in the
different spectrum of operators with nonzero $W$ in these two
compactifications.

The spectrum of closed string vertex operators in the thermal THO
theory is \Davisup
 \eqn\thotherm{\eqalign{
  &T_C=e^{-\varphi}\bar \CO_C(\half +2nw) V_{n,2w}
  \cr
  &T_0=e^{-\varphi}\bar \CO_0(\half +(n+\half)(2w+1)) V_{n+\half,2w+1}
  \cr
  &\Psi_S = e^{-\half\varphi + i\half H }\bar \CO_S(
  \half +(2n+1)w) V_{n+\half,2w}~,\qquad ~~p_R \geq 0  \cr
  &\Psi_V = e^{-\half\varphi + i\half H }\bar \CO_V(
  \half +n(2w+1)) V_{n,2w+1}~,\qquad ~~p_R \geq 0  \cr
  &\tilde \Psi_S = e^{-\half\varphi - i\half H }\bar \CO_S(
  \half +n(2w+1)) V_{n,2w+1}~, \qquad ~~p_R \leq 0  \cr
  &\tilde \Psi_V = e^{-\half\varphi - i\half H }
  \bar \CO_V(\half +(2n+1)w) V_{n+\half,2w}~, \qquad ~p_R \leq 0 \cr
  }}
This spectrum is invariant under $R \to {1\over R}$ up to
spacetime symmetry.  This T-duality is not guaranteed by an
enhanced non-Abelian symmetry at the selfdual point.

The closed strings lead to \refs{\DavisQE,\Davisup}
 \eqn\gammath{\eqalign{
 \log\left(\CZ_{ closed}\right) = & V
 \Gamma_{ closed}+\CO(1) \cr
 \Gamma_{ closed}  = & \cases{
 {1\over R}  & HE \cr
 R+{2\over R}& HO \cr
 -{R\over 2}+{1\over 2 R} & THO }}}
and the worldsheet answers are \refs{\DavisQE,\Davisup}
 \eqn\gammathr{ \eqalign{
 \Gamma_{worldsheet}(R>1) =&\Gamma_{closed}(R>1) \cr
 \Gamma_{worldsheet}(R<1) = &\Gamma_{closed}(R<1)
 +\left(R-{1\over R}\right)
 =  \cases{
 R  & HE \cr
 2R+{1\over R}& HO \cr
 {R\over 2}-{1\over 2 R} & THO }}}
where we see the effect of the massless boson at $R=1$.

Repeating the analysis of the charges of the lowest energy state
in the $\phi$-channel, we find that only in the THO theory the
ground state needs a modification, and that is achieved by acting
on it with $T_0(N=-\half, W=1)$.  Therefore,
$\Gamma=\Gamma_{worldsheet} - \left|{R\over 2} - {1\over
2R}\right| $. We conclude that
 \eqn\gammathg{ \eqalign{
 \Gamma (R>1)  =& \cases{
 {1\over R}  & HE \cr
 R+{2\over R}& HO \cr
 -{R\over 2}+{1\over 2 R} -\left({R\over 2} -{1\over 2R}\right) =
 -R+{1\over R} & THO } \cr
 \Gamma (R<1)  = &\cases{
  \Gamma_{worldsheet}(R<1) = R  & HE \cr
 \Gamma_{worldsheet}(R<1) = 2R+{1\over R}& HO \cr
 \Gamma_{worldsheet}(R<1) - \left|{R\over 2} - {1\over
 2R}\right|=R-{1\over R} & THO }}}
As expected, these results T-dual.  Note that the added string in
THO with $W=1$ and $N=-\half$ becomes after T-duality a string
with $W=-1$ and $N=\half$.  Its opposite orientation fits with the
fact that the T-dual theory is the parity image of THO.

\newsec{Conclusions}

We have studied the two-dimensional heterotic string uncovering a
number of new phenomena:

\item{1.} The THO theory has anomalies which are cancelled by
adding a single infinitely long string stretched across the space.

\item{2.} The spectrum of all our theories includes pairs of a
long string and a long anti-string, or equivalently, the folded
string of figure 2. These infinite energy excitations are
important in the study of compactifications.

\item{3.} After compactification of the $x$ direction we find a
generalization of the need to add a long string.  In some of our
examples a careful analysis of the gauge symmetries of the
compactified theory shows that the lowest energy state in the
$\phi$-channel is not invariant.  This lack of gauge invariance
can be fixed by acting on it with a creation operator.  This has
the effect of changing the worldsheet answer.  The latter always
describes the lowest energy state whether it is invariant or not.

\item{4.} As we vary the parameters of the compactifications
including the radius $R$ and Wilson lines various phase
transitions can take place.  The simplest kind of phase transition
is associated with massless fermions.  Since the fermions are
charged, the lowest energy states in the $\phi$-channel in the two
sides of the transition have different quantum numbers. Therefore,
at least on one side of the transition (and possibly on both
sides) we need to act on the lowest energy state with a vertex
operator. This changes the transition and sometimes smoothes it
out.

\item{5.} Our most dramatic conclusions are associated with phase
transitions associated with massless bosons.  These arise, among
other places, in thermal compactifications.  It seems that we have
a conflict between T-duality, locality in $\phi$, locality in $x$
and locality in the dual of $x$.  One possibility is to follow the
quantization in the $\phi$-channel and preserve T-duality.  Then
we have standard thermodynamics in the $x$-channel for $R>1$ (low
temperatures) and standard thermodynamics in the dual of $x$ for
$R<1$ (which is again low temperatures, in the T-dual picture).
But the $x$-channel thermodynamics cannot be continued to higher
temperatures.

We would like to make a few more comments about the last point
regarding the thermal circle.

Let us compare our thermal system with the thermal ten-dimensional
heterotic string.  In both cases we find a thermal phase
transition due to the winding mode with $N=-\half$, $W=1$ and its
complex conjugate with $N=\half$, $W=-1$.  The corresponding
worldsheet vertex operator is relevant for $\sqrt 2 -1< R <1+\sqrt
2$.  In the critical string it is massless at $R=\sqrt 2 \pm 1$,
and it is tachyonic between these two points. This is the standard
Hagedorn transition.  In the two-dimensional heterotic string the
vertex operator is still relevant for $\sqrt 2 -1< R <1+\sqrt 2$,
but it is never tachyonic.  It is massless only at the selfdual
point $R=1$. Therefore, we expect the transition in two dimensions
to be milder than in ten dimensions.

We assume that these heterotic theories have dual descriptions in
terms of boundary theories.  What do our results imply about these
holographic descriptions? The standard AdS/CFT \AharonyTI, and the
$c\le 1$ and $\hat c\le 1$ theories have a dual description in
terms of a local quantum field theory. Even though our heterotic
theories have a large dilaton slope and only a finite number of
massless particles in the two-dimensional bulk, they are similar
to higher dimensional linear dilaton theories like the little
string theory \refs{\BerkoozCQ\SeibergZK\AharonyUB-\AharonyKS}.
These holographic theories do not appear to be local
\refs{\SeibergZK,\PeetWN,\AharonyXN}, and they have peculiar
finite temperature behavior \AharonyXN.

One might ask wether our analysis, which focuses only on the bulk
of the target space and ignores the strong coupling region, can be
misleading.  In particular, can the system have instabilities
which are not visible in the bulk but are supported entirely in
the strong coupling region?  Examples of such instabilities were
recently discussed in \ItzhakiZR.   However, the ability to turn
on the deformation \mudef\ suggests that this is not the case.  As
we commented above, this deformation prevents the strings from
penetrating into the strong coupling region.  Furthermore, unlike
the examples of \ItzhakiZR, where the instability is associated
with localized tachyons, it is clear that with \mudef\ no
localized tachyons can be present.

We can study an alternative quantization in the $\phi$-channel. We
considered the ground state wave function $\psi_0 \sim
e^{-{|m|\over 2} |\Phi|^2}$. Alternatively, we might want to
consider the wave function $\tilde \psi_0 \sim e^{-{m\over 2}
|\Phi|^2}$.  It is the same as $\psi_0$ for positive $m$, but it
is not normalizable for negative $m$. This wave function can be
interpreted as associated with a state $|\tilde 0\rangle$ which is
annihilated by the creation operators of the theory.  Its energy
is $-|m|$ and acting on it with $U(1)$ invariant combinations of
annihilation operators leads to states $|\tilde n\rangle$ with
energy $-|m|(2\tilde n+1)$ ($\tilde n=0,1,...$). This quantization
of the harmonic oscillator leads to a spectrum which is unbounded
from below and is also not unitary. Therefore, it is usually
ignored. If we follow this strange quantization in the
$\phi$-channel, we find answers which are compatible with the
$x$-channel picture. One way to understand it is to note that here
the energy of $|\tilde 0\rangle $ is $m$ rather than $|m|$, and
the annihilation operators which lead to lower energy states have
energy $-\Gamma_{long} = m={R\over 2} -{1\over 2R}$ without
absolute value.  Then, as for positive $m$ the state $|\tilde
n\rangle$ of the $\phi$-channel can be interpreted in the
$x$-channel as having $\tilde n$ pairs. Unfortunately, the
partition function associated with that quantization diverges.
Also, this quantization treats differently positive and negative
$m$, and therefore it does not respect the T-duality of our
system.

Finally, we would like to mention some logical options which could
invalidate our conclusion about the lack of locality in the
$x$-channel evolution. It might be that our theories are simply
inconsistent. Alternatively, they are inconsistent only for some
range or $R$ which includes $R=1$. Finally, perhaps $R=1$ should
be viewed as a limiting radius and a limiting temperature, and the
Euclidean circle cannot be reduced below $R=1$.  Clearly, this
last possibility is inconsistent with T-duality.

\bigskip
 \centerline{\bf Acknowledgements}
It is a pleasure to thank J.L.~Davis for collaboration during the
early stages of this project. We have also benefited from useful
discussions with O.~Aharony, N.~Itzhaki, I.~Klebanov, D.~Kutasov,
F.~Larsen, H.~Liu, J.~Maldacena and E.~Witten. This work was
supported in part by grant \#DE-FG02-90ER40542.

\appendix{A}{Type 0/II}

In this appendix we will reexamine the various compactifications
of the type 0 and type II theory of \SeibergBX\ and check whether
a modification of the $\phi$-channel ground state is needed.

In \SeibergBX\ a moduli space consisting of eight lines of
theories was identified.  We find that five of the lines do not
need any modification.  However, in the three lines corresponding
to compactifications of the type IIA theory, the naive
$\phi$-channel vacuum is not invariant. Using the notation of
\SeibergBX\ we find:

\item{Line 4}  This is a circle compactification of IIA on a
circle of radius $R$.  It is dual to IIB on a circle of radius
$R'={2\over R}$.  In the IIB language we have a complex periodic
fermion of one chirality and a periodic scalar with the opposite
chirality. Here the naive ground state has $N= {1\over 8}$.  This
translates in IIA to $W={1\over 8}$. Therefore we might need to
change $\Gamma \to \Gamma - {R \over 16}$.

\item{Line 6}  This is the superaffine IIA theory, and it is
selfdual under $R \to {1\over R}$. Here we have an antiperiodic
fermion of one chirality and a periodic fermion with the opposite
chirality. Therefore, the naive ground state has $N={1\over 16}$,
and using T-duality, $W={1\over 8}$. Therefore we might need to
change $\Gamma \to \Gamma - {R \over 16}- {1\over 16 R}$.

\item{Line 8}  This is a compactification of IIA on a thermal
circle of radius $R$.  It is dual to 0B on a twisted circle of
radius\foot{This corrects a misprint in \SeibergBX.} $R'={1\over
R}$.  In the 0B language the left-moving and the right-moving
tachyons are anti-periodic. One chirality of the $C$ field is
periodic and the other chirality is antiperiodic. Therefore, the
naive ground state has $N={1\over 16}$, or in the IIA language
$W={1\over 8}$, and we might need to change $\Gamma \to \Gamma -
{R \over 16}$.

These modifications of the ground state are puzzling for two
reasons.  First, the theory does not have closed string vertex
operators with these quantum numbers.  Second, it suggests a
modification of the IIA theory with noncompact $x$ by an insertion
with nonzero $W$. Such an insertion violates the parity symmetry
of the theory.

We do not have a clear understanding of this puzzle.  Perhaps
these compactifications are inconsistent.  Alternatively, perhaps
for some reason no modification is needed here.  The most likely
possibility is the following.  These theories have excitations
which cannot be described by worldsheet vertex operators.  Such
excitations of the type 0 and type II theories were discussed in
\refs{\DeWolfeQF,\SeibergBX,\MaldacenaHE}. Perhaps these
excitations can be used to achieve the invariance of the vacuum.

\listrefs
\end